\documentclass[11pt,a4paper]{article}
\pdfoutput=1
\usepackage[utf8]{inputenc}
\usepackage[english]{babel}

\usepackage{extarrows}
\usepackage{amsmath,bm}
\usepackage{amsfonts}
\usepackage{amssymb}
\usepackage{graphicx}
\usepackage{fourier}

\usepackage{empheq}

\usepackage{caption}
\usepackage{subcaption}
\usepackage{float}
\usepackage{appendix}
\usepackage{authblk}

\numberwithin{equation}{section}

\usepackage[hidelinks]{hyperref}

\newcommand\nin{\noindent}
\newcommand\nn{\nonumber}
\newcommand\be{\begin{equation}}
\newcommand\ee{\end{equation}}
\newcommand\ba{\begin{eqnarray}}    
\newcommand\ea{\end{eqnarray}}      

\title{Holographic excited states in AdS Black Holes}
\author{Marcelo Botta-Cantcheff}
\author{Pedro J. Mart\'inez}
\author{Guillermo A. Silva}
\affil{{\it  Instituto de F\'\i sica de La Plata, CCT La Plata - CONICET \& 

Departamento de F\'\i sica, Universidad Nacional de La Plata  

C.C. 67, 1900 La Plata, Argentina}

~

{\tt E-mail:} botta,martinezp,silva@fisica.unlp.edu.ar}

\usepackage[left=2cm,right=2cm,top=2cm,bottom=2cm]{geometry}

\begin{document}
\maketitle
\begin{abstract}
We have recently presented a geometry dual to a Schwinger-Keldysh  closed time contour, with two equal $\beta/2$ length Euclidean sections, which can be thought of as dual to the Thermo Field Dynamics formulation of the boundary CFT.  In this work we study non-perturbative holographic excitations of the thermal vacuum by turning on asymptotic Euclidean sources. In the large-$N$ approximation the states are found to be thermal coherent states and we manage to compute its eigenvalues. We pay special attention to the high temperature regime where the manifold is built from pieces of Euclidean and Lorentzian black hole geometries. In this case, the real time segments of the Schwinger-Keldysh contour get connected by an Einstein-Rosen wormhole through the bulk, which we identify as the exterior of a single maximally extended  black hole. The Thermal-AdS case is also considered but, the Lorentzian regions become disconnected, its results mostly follows from the zero temperature case.
\end{abstract}

\tableofcontents

\newpage
\section{Introduction}

AdS/CFT \cite{adscft} is mostly developed in Euclidean time \cite{GKP,W}. Conceptually, there is no fundamental principle forcing an Euclidean formulation of the duality.  
However, a direct approach to real time holography give raise to subtleties \cite{Marolf:Excited-States}.
In particular, real time evolution demands initial and final conditions which are not immediate to characterize from both sides of the duality, in conflict with a strict holographic viewpoint.  

The Skenderis and van Rees (SvR) prescription \cite{SvRC,SvRL} provides a completely holographic real time extension of the GKPW standard prescription \cite{GKP,W}. It essentially maps the initial/final state information, through auxiliary Euclidean regions, to boundary data in the CFT. The general set-up of the SvR prescription thus deals with manifolds of mixed signature, the philosophy being to require only holographic/boundary data. 

In the SvR framework, sources on the Lorentzian asymptotic boundary are thought of as devices to obtain $n$-point correlation functions.
On the other hand, Euclidean sources play a very different role and prepare the state of the system at a given time. The foundational works \cite{SvRC,SvRL}, for example, showed that turned off Euclidean sources prepare the vacuum state.  
Turned on sources, in turn, allow to prepare (holographic) excited states of the CFT \cite{us}, see also \cite{Ariana} for related work.

In \cite{us} we began the study of general non-trivial sources in manifolds with mixed signature complementing the bulk treatment with the BDHM dictionary \cite{BDHM}. With this machinery we were able to show that, in the large N approximation, the excited states obtained by turning on Euclidean sources are coherent states. Interacting fields in the bulk lead to states of a modified nature, which we analyzed in \cite{us2}. A more systematic understanding of these excitations is under development. These holographic excited states have been of interest in recent literature \cite{Aitor-EntanglementyEinsteinEq}-\cite{Belin:2018bpg}.

In a recent work \cite{prev}, we presented a novel geometry dual to a Schwinger-Keldysh (SK) contour \cite{Schwinger} describing real time evolution of a finite temperature CFT in which standard Thermo Field Dynamics (TFD) \cite{UmezawaAFT} computations can be carried holographically. We studied the geometry, its two-point functions and its role in the context of the Hawking-Page (HP) transition. For comparison, in this same work we studied the real time extension of Thermal-AdS. The main objective of the present work is to study holographic excited states on these finite temperature geometries. 

We will provide a review of the formalism developed in \cite{us} and derive its extension to the finite temperature set-up. The most relevant result of \cite{us} that we will exploit in the present paper is the In-Out formulation, that allowed to split and interpret  the  Euclidean and Lorentzian path integral pieces as initial/final states and real time evolution of the system respectively. This splitting permitted us 
to study of the excited states as objects (kets) independently of the precise SK path it is glued to, e.g. a semi-infinite Euclidean path integral with non-zero sources corresponded to a precise holographic state, coherent in the large-N limit. In this work we pursue an analogous objective for the geometry we built in \cite{prev}. Its TFD interpretation will provide the required In-Out structure. Previous thermal geometries \cite{SvRL,eternal,Balt} were not suitable for this interpretation.

We will compute inner products and matrix elements of CFT local operators for holographic excited states, the latter directly related to linear response quantities in standard TFD formalism. The inner products, which require collapsing the real time segments, can be understood as a reinterpretation of standard Euclidean result with non-zero sources. The kernels in these objects, due to the coherent nature of the excited states, define Kähler potential in the space of states which may be of interest for the developments in \cite{Aitor-Simplectic}. The matrix elements on their own help to recognize the thermal coherent character of the states and to  determine its eigenvalues.

The path integral approach demands finding a general solution to the equations of motion with non-trivial Euclidean and Lorentzian sources turned on. We will build it in detail, checking that CFT information is enough to give a unique analytic solution inside the bulk. This result is non-trivial once we notice that the Lorentzian Rindler-like patches, dual to real time evolutions, end up being glued analytically through an Einstein-Rosen (ER) wormhole. We will show that this property relies on the Euclidean sections of the SK path having identical extensions.
 
We will also work with the BDHM formalism at finite temperature, this is the natural framework both to demonstrate the coherent nature and to compute the eigenvalues of the holographic excited states.  The bulk field will be quantized on a patch covering only the exterior of a maximally extended AdS black hole (AdSBH) and Unruh-like global modes over the wormhole will be obtained by demanding analyticity on the radial ER coordinate. In the end, the  excitations over the TFD vacuum state turn out to be \textit{thermal} coherent states \cite{Thermal-Coherent}.

The paper is organized as follows. In Sec. \ref{SK+CFT-2nd} we review the SvR prescription in a general path integral formulation and the rules to construct a TFD double. Specifically, Sec. \ref{Sec:ClosedPaths} contains new results: excited states are constructed in terms of {\it evolution operators} on the Euclidean pieces of a symmetric closed SK contour, whereas the evolution operators that arise from the two real segments in the complex $t$-plane are combined and identified with that of the TFD double, allowing to interpret the field theory SK path integral as In-Out scattering process at finite temperature\footnote{To avoid confusion $T$ will always refer to time. Temperature will solely be denoted by $\beta$.}. In Sec. \ref{Sec:Bulk}  excited states are studied from the path integral SvR approach in the semi-classical limit. We pay particular attention to the construction of the bulk field solution with general sources and study its analytical properties. We then compute inner products between our states and matrix elements of boundary local operators. This will make manifest  the fact that initial/final excitations are not defined in a single Hilbert space but rather in a doubled space. The details of the computations for the low temperature regime, i.e.  Thermal AdS, are relegated to App. \ref{App:Thermal}. In Sec. \ref{Sec:BDHM} we complement  the study of the states incorporating the BDHM viewpoint into the analysis. This allows to precisely identify them as coherent in terms of the Bogoliubov rotated operators, predicting also its eigenvalues.
Finally, Sec. \ref{Sec:Conclusion} summarizes the results and discusses possible future applications. 

\section{The SvR approach, excited states, and In-Out formalism}
\label{SK+CFT-2nd}

 In this section we define holographic excited states  at finite temperature and focus  on its field theory description. We start by reviewing the SvR construction \cite{SvRC} and its extension, developed in \cite{us}, to consider excited states in open complex contours. The open path scenario motivates a splitting of the standard GKPW formula into a piecewise holographic prescription. We then clarify some aspects of our results concerning the definition of the excited states. Using TFD language, we then reinterpret closed Schwinger-Keldysh paths as finite temperature scattering processes and characterize the structure of the holographic excited states using the TFD formalism. Finally, we write a piecewise holographic map which links Euclidean sections to initial/final excited states and Lorentzian ones to (boost-like) time evolution in the black hole (BH) geometry.

\subsection{Brief review of the In-Out formalism: open paths}

\begin{figure}[t]\centering
\begin{subfigure}{0.49\textwidth}\centering
\includegraphics[width=.9\linewidth] {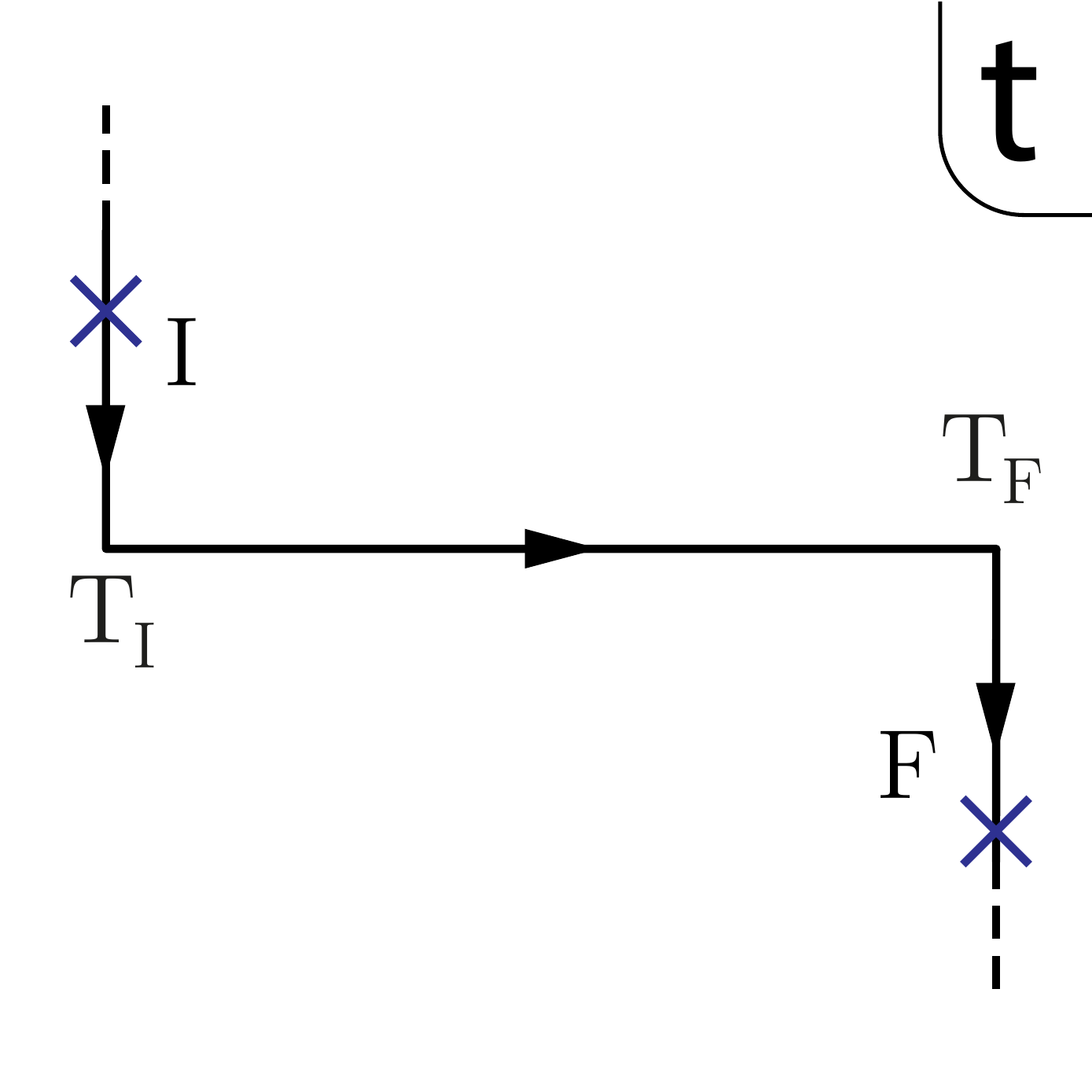}
\caption{}
\end{subfigure}
\begin{subfigure}{0.49\textwidth}\centering
\includegraphics[width=.9\linewidth] {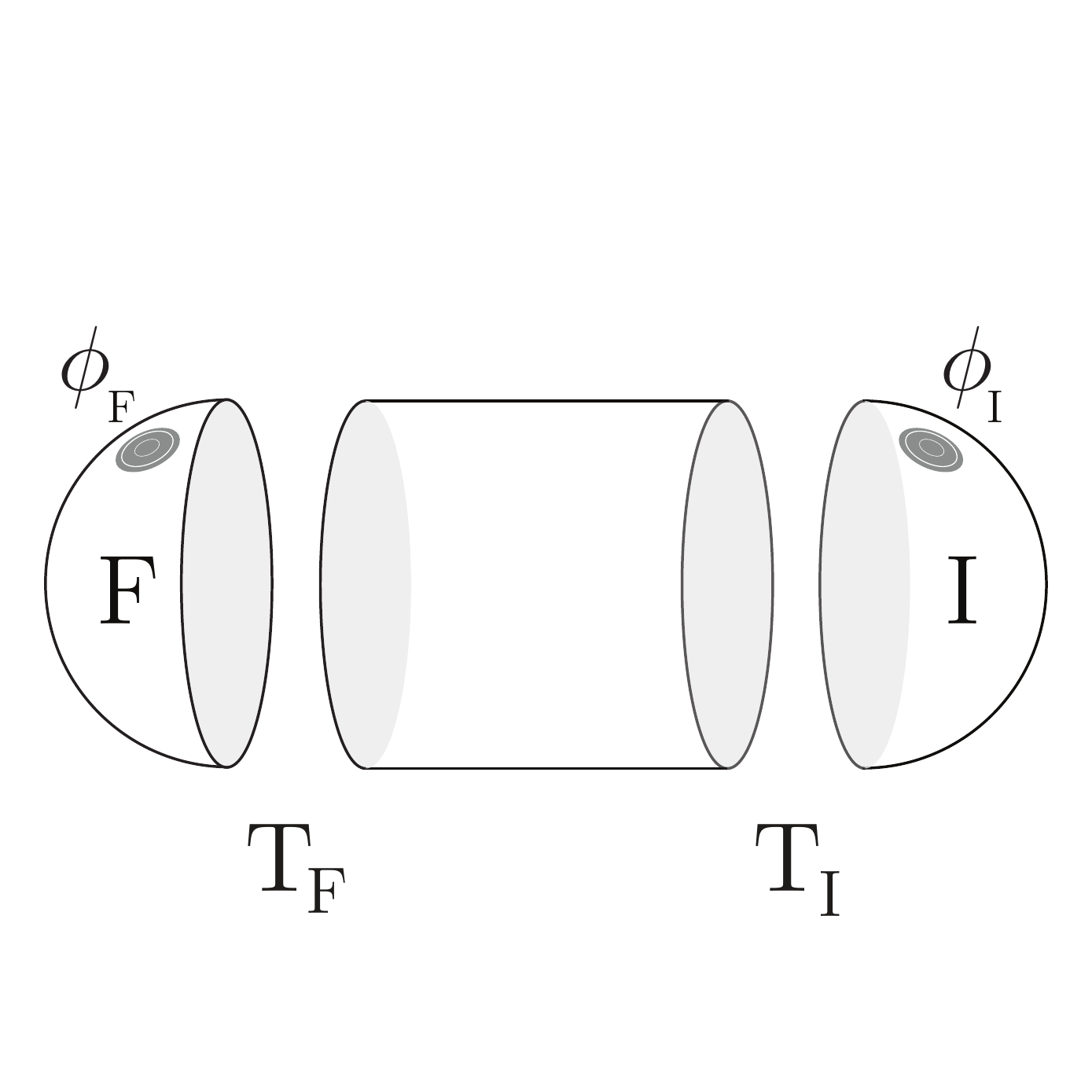}
\caption{}
\end{subfigure}
\caption{(a) In-Out SK path representing a scattering experiment. The (horizontal) Lorentzian piece describes the system evolution while the (vertical) Euclidean I/F pieces, with blue crosses denoting local operator insertions, prepare excited initial and final states at times $T_{I/F}$. (b)  The semi-classical bulk dual description of the same problem fills the time contour with Euclidean and Lorentzian AdS manifolds. Sources on the Euclidean segments generate holographic excited states.}
\label{Fig:Pastilla}
\end{figure}

The SvR holographic prescription can be summarized in the following formula
\begin{equation}\label{SvRPath}
Z_{CFT} \left[\phi({\cal C})\right]
=
Z_{grav}\left[\Phi|_{\partial}=\phi\left({\cal C}\right)\right]
\end{equation}
where the lhs is the generating function for correlation functions of CFT operators ${\cal O}$ with the sources $\phi({\cal C})$ having support on a specific continuous path ${\cal C}$ in the complex $t$-plane. The rhs is the partition function for the bulk field $\Phi$, dual to $\cal O$, on an aAdS spacetime with asymptotic boundary conditions $\phi({\cal C})$.
This general path integral expression applies to all  contours ${\cal C}$, concomitantly the dual spacetimes combine both signatures \cite{SvRC,SvRL}, and in particular reduces to the purely Euclidean set up \cite{GKP,W} as the real-time intervals are removed, or Wick-rotated.

In the so-called In-Out formalism, investigated in \cite{SvRC, SvRL}, one  considers \emph{open} contours, let us refer to them as ${\cal C}_O$. The  curve ${\cal C}_O \equiv \{  t-i\tau \in \mathbb C\} = {\cal C}_I \cup {\cal C}_L \cup {\cal C}_F$ in the time complex plane is divided in three pieces as depicted in Fig. \ref{Fig:Pastilla}(a),
 and the path ordering ${\cal P}$ follows the arrowed lines in each subset. 
Then, the prescription above takes the  explicit form
\begin{equation}
\label{wavef-g}
\langle 0 | \, U \,|0 \rangle = \int_{\Phi|_{\partial}=\phi({\cal C}_O)} [\mathcal{D}\Phi] \,  e^{i S^{}[\Phi]}\,\, , 
\end{equation}
where  $U$ is the evolution operator given by the  CFT Hamiltonian deformed with a single-trace operator ${\cal O} \equiv {\cal O}({\bf x})$, multiplied by an arbitrary  time dependent source\footnote{To simplify notation, the ${\bf x}$-integration is left implicit.} $\phi({\bf x}, \theta)$:
\begin{equation}\label{ZCFT-open}
 U \equiv {\cal P} \,e^{-i\int_{{\cal C}_{O}} d\theta\; (H + {\cal O} \,\phi(\theta))}\,
\end{equation}
The state $|0 \rangle$ is the CFT vacuum  expressed in the Schr\"odinger picture and the curve ${\cal C}_{O}$ is parameterized  so that $d\theta= -id\tau$ on ${\cal C}_{I,F}$, and $d\theta=dt$ on ${\cal C}_{L}$.  

In its original form, the above proposal was studied in the semi-classical limit of the gravitational side, which corresponds to the large N limit in the standard AdS/CFT example \cite{SvRC}:
\be\label{SvRpath-open + large N}
Z_{CFT}[\phi({\cal C}_O)] =\langle 0 | \, {\cal P} \,e^{-i\int_{L} d t\; (H + {\cal O} \,\phi(t))} \,|0 \rangle \approx e^{-i S^0[\phi({\cal C}_O)]} \, 
\ee
with the boundary conditions $ \Phi|_{{\cal C}_{I,F}}\equiv 0$, although, it was also claimed that by imposing non vanishing asymptotic boundary conditions in the Euclidean regions $\phi_{I,F} = \Phi|_{{\cal C}_{I,F}}\neq 0$, this formula should generalize to account for excited in/out states. 
This statement was explicitly verified in  \cite{us}, splitting (\ref{ZCFT-open}) as $U = U_F \, U_L \, U_I $ one gets an explicit formula for the holographic excitations
\begin{equation}\label{state-open}
 |\Psi_{\phi}\rangle =  U_{\phi} |0\rangle= {\cal P} \,e^{-\int_{\tau<0} d\tau \; (H + {\cal O} \,\phi(\tau))}|0\rangle
\end{equation}
 which become parametrized by the arbitrary source $\phi ({\bf x}, \tau) $ with compact support on the interval $\tau \in (-\infty, 0)$ \cite{us}.
In the interaction picture this state can be written as
\begin{equation}\label{state-open-interaction}
 |\Psi_{\phi}\rangle =  \,{\cal P}\,e^{-\int_{\tau<0} d\tau \; {\cal O}({\bf x}, \tau) \,\phi({\bf x}, \tau)}|0\rangle 
\end{equation}
where  ${\cal O}({\bf x}, \tau )\equiv e^{\tau H} {\cal O}({\bf x}) e^{-\tau H} $, and $|0\rangle_{I}= e^{-\tau H}|0\rangle = |0\rangle$.
The corresponding duals (``\textit{bra}") of these \emph{kets}, are built by taking the Hermitian conjugate of the Euclidean evolution operator: $U_{\phi} \to U_{\phi}^\dagger \,\equiv\, U^{}_{\phi^*}$ in (\ref{state-open}). 
This  operation defines 
the source $\phi^* \equiv \phi ({\bf x}, -\tau)\;, \tau \in (-\infty,0)$ on
 the interval $\tau \in (0, \infty)$, see \cite{us,Jackiw}.
Thus, in the interaction picture  reads 
\begin{equation}\label{state-open-interaction-bra}
 \langle\Psi_{\phi} | = \, \langle 0|\,{\cal P}\,e^{-\int_{\tau > 0} d\tau \; {\cal O}^\dagger({\bf x}, \tau) \,\phi^{*}({\bf x}, \tau)}\;.
\end{equation}
It has been stressed that states of this form are holographic in the sense that correspond to  well defined geometric duals \cite{Faulkner2017,Marolf,Botta-Cantcheff:2017gys,Mosk}.  

From the bulk perspective, one can consider a co-dimension one, spacelike hypersurface $\Sigma$ in the bulk, whose boundary intersects the contour ${\cal C}_O$ at the point $\tau =0$ ($\partial \Sigma = S^d $). An arbitrary {\it initial} data $\phi_\Sigma := \{ \phi(x) \;, \; x\in \Sigma \}$ , representing the eigenvalue of the quantized bulk local field operator $\widehat{\Phi}(x)$, can be inserted in both sides of the path integral \eqref{wavef-g} at $\tau =0$, and then summing over, one obtains a piece-wise version of this prescription, see \cite{us} for more details.
The state \eqref{state-open} projected on this basis provides the gravity wave functional on the right hand side \cite{SvRC,SvRL,us},
\begin{equation}
\langle \phi_{\Sigma} |\, U_{\phi} |0\rangle \equiv\langle\phi_{\Sigma}|\Psi_\phi\rangle = \Psi_{\phi}(\phi_{\Sigma}) = \int^{}_{}[{\cal D}\Phi]_{(\phi_{\Sigma} , \,\phi)} \,\; e^{-S_E[\Phi]}\;, 
\end{equation}
which, by virtue of the asymptotic boundary condition (source) $\phi \neq 0$, generalizes the Hartle-Hawking quantum gravity wave functionals, to excited states \cite{HH,us}. Recall that this path integral implicitly includes the (formal) sum over the gravitational degrees of freedoms. 

One of the most interesting features of \eqref{state-open-interaction} is that, by canonically quantizing a (nearly) free non-back reacting field $\Phi$ in the bulk, these states become coherent in the large $N$ Hilbert space \cite{us}  
\begin{equation}\label{coherent-state}
|\Psi_{\phi}\rangle \propto e^{\int dk\; \lambda_k a^\dagger_k}|0\rangle\,.
\end{equation}
Here $a_k\,( a^\dagger_k\,)$ are the annihilation (creation) operators associated to  the canonically quantized bulk field $\hat{\Phi}$ and $\lambda^{}_k$ are eigenvalues of $a_k$, given by the Laplace transform of the Euclidean sources. This result is obtained by using the so-called BDHM prescription that relates CFT local operators with quantized bulk field operators, i.e, the operators ${\cal O}$ of (\ref{state-open-interaction}) are linearly expanded in terms of $a_k\, , \, a^\dagger_k$, see \cite{BDHM,kaplan,us}.

Below, by working with the TFD formalism, we will see how the In-Out formalism can be extended to the case of closed paths 

\subsection{Closed paths: the Schwinger-Keldysh contour and TFD}\label{Sec:ClosedPaths}

\begin{figure}[t]\centering
\begin{subfigure}{0.49\textwidth}\centering
\includegraphics[width=.9\linewidth] {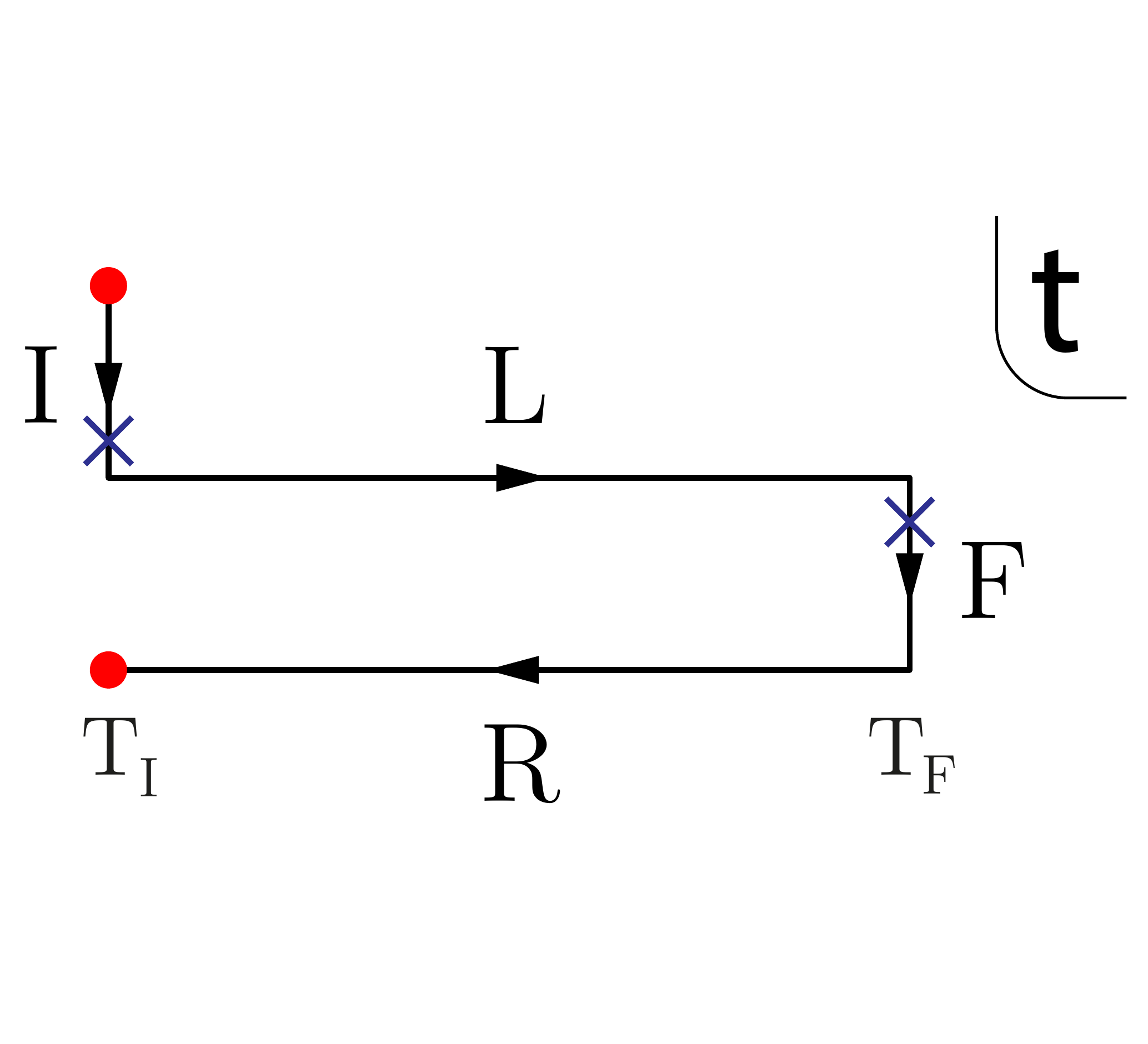}
\caption{}
\end{subfigure}
\begin{subfigure}{0.49\textwidth}\centering
\includegraphics[width=.9\linewidth] {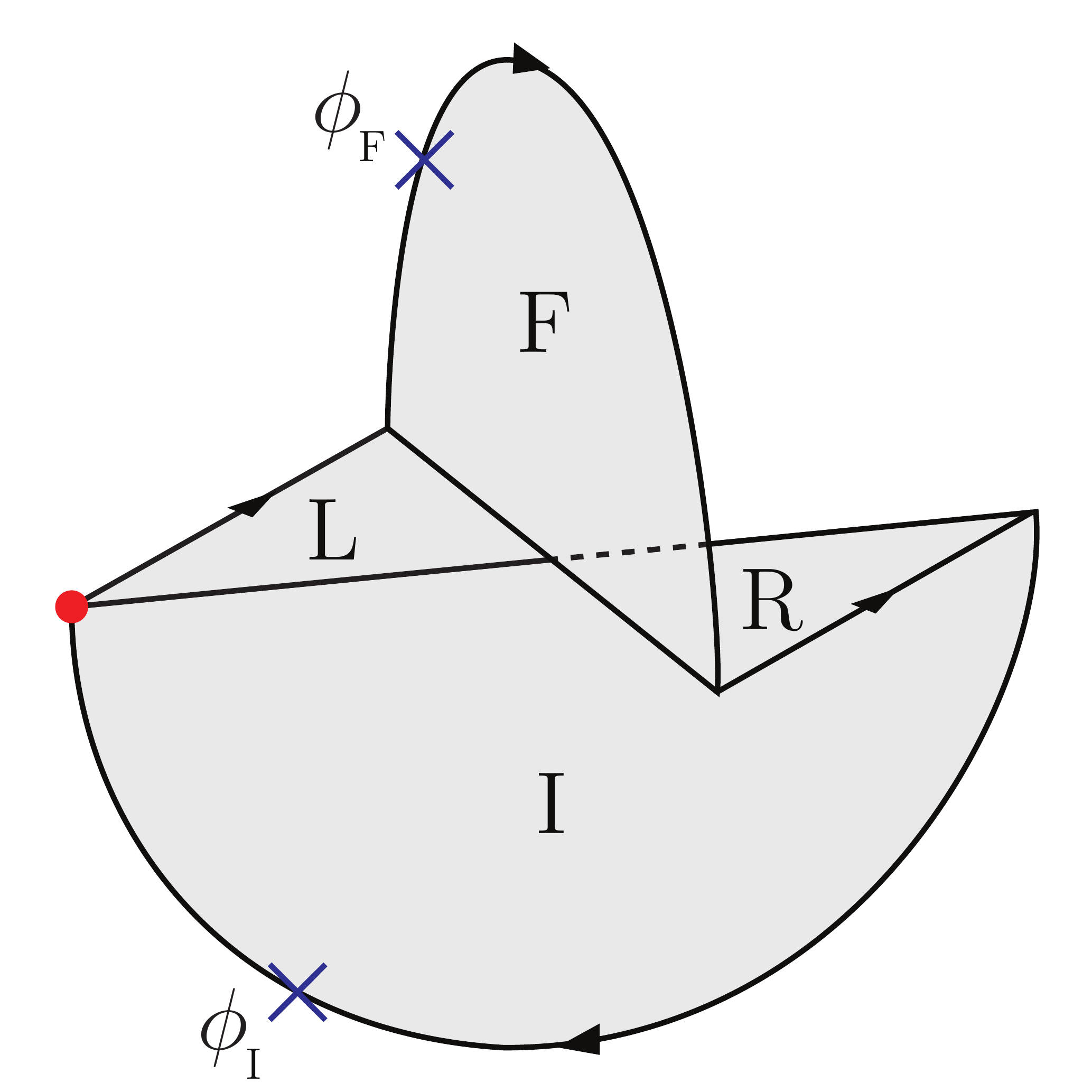}
\caption{}
\end{subfigure}
\caption{(a) Closed Schwinger-Keldysh path in the complex $t$-plane.  The horizontal lines represent real time evolution. The vertical lines give imaginary time evolution, and the regions I and F have identical lengths equal to $\beta/2$.  The insertion of sources in the vertical lines generate excitations over the (vacuum) thermal state. (b) Dual bulk geometry filling the path in fig. (a): the semicircular pieces represent the Euclidean sections. The horizontal plane depicts the (static) AdSBH exterior represented as triangular wedges L and R. The angular coordinate has been suppressed.}
\label{Fig:Camino}
\end{figure}

 Let us now apply the construction discussed above to a CFT defined on a \emph{closed} time contour ${\cal C}$ in the complex plane. The symmetric Schwinger-Keldysh path involving two imaginary-time intervals (of  equal length $\beta/2$) \cite{Schwinger,Umezawa:TFD=beta/2}, shown in Fig. \ref{Fig:Camino}(a), was investigated in \cite{prev} in the holographic context. 

For a closed path, the lhs of \eqref{SvRPath} is expressed as follows 
\begin{equation}\label{ZCFT}
Z_{CFT} = \text{Tr} \, \;U \qquad\qquad U \equiv {\cal P} \,e^{-i\int_{{\cal C}_{}} d\theta\; (H + {\cal O} \,\phi(\theta))} 
\end{equation}
The evolution operator in this case factorizes as $U = U_R U_{F} U_L  U_{I}$, where $U_{L/R}$ are ordinary real time evolution operators with the CFT Hamiltonian $H$ deformed by external (local) sources $\phi(\textbf{x},t)$. The operators $U_{I,F}$ on both imaginary time intervals, univocally describe the initial/final excited states in terms of the local sources $\phi(\textbf{x},\tau)$ \cite{prev}:
\begin{equation}\label{Upm}
 U_{\phi} \equiv {\cal P} \,e^{-\int d\tau \; (H + {\cal O} \,\phi(\tau))}\,. 
\end{equation}
where the Euclidean time $\tau$ runs on the intervals $(-\beta/2, 0)$ and $(0,\beta/2)$ on I and F respectively.
In fact, in the TFD context these operators are equivalent to pure states, rearranged as \textit{kets} in the duplicated states space \cite{prev}.

Let us consider a conformal field theory, whose states belong to the Hilbert space ${\cal H}$. 
In the TFD formalism, one constructs a second copy of the system, namely $\widetilde{{\cal H}}$, so that the total new system consist of the original CFT and its TFD copy, living on disconnected asymptotic boundaries of the gravity dual, with ${\cal H}\otimes\widetilde{{\cal H}}$ total states space  \cite{UmezawaAFT,Takashi-TFD}. Thus, for any given operator $A$, acting on $\cal H$, one builds the corresponding operator $\widetilde{A}$  on $\widetilde{{\cal H}}$ using the so-called ``tilde'' conjugating map \cite{UmezawaAFT,Thermal-Coherent},
\begin{align}\label{Tilde-rules}
[A,\tilde B]=0 && (AB)\tilde{\,}=\tilde A \tilde B &&
(c_1 A + c_2 B)\tilde{\,} = c_1^* \tilde A + c_2^* \tilde B && (A^{\dagger})\tilde{\,}=\tilde A ^{\dagger}\;.
\end{align}

Alternatively, one can denote the extended operators as $A_L$ and $A_R$ respectively: 
\be\label{R-tilde} A\otimes\mathbb{1} =: A_L \;,\qquad\qquad \tilde (A\otimes \mathbb{1}) =\mathbb{1}\otimes \tilde{A}  =: A_R\;,
\ee

\begin{figure}[t]\centering
\begin{subfigure}{0.49\textwidth}\centering
\includegraphics[width=.80\linewidth] {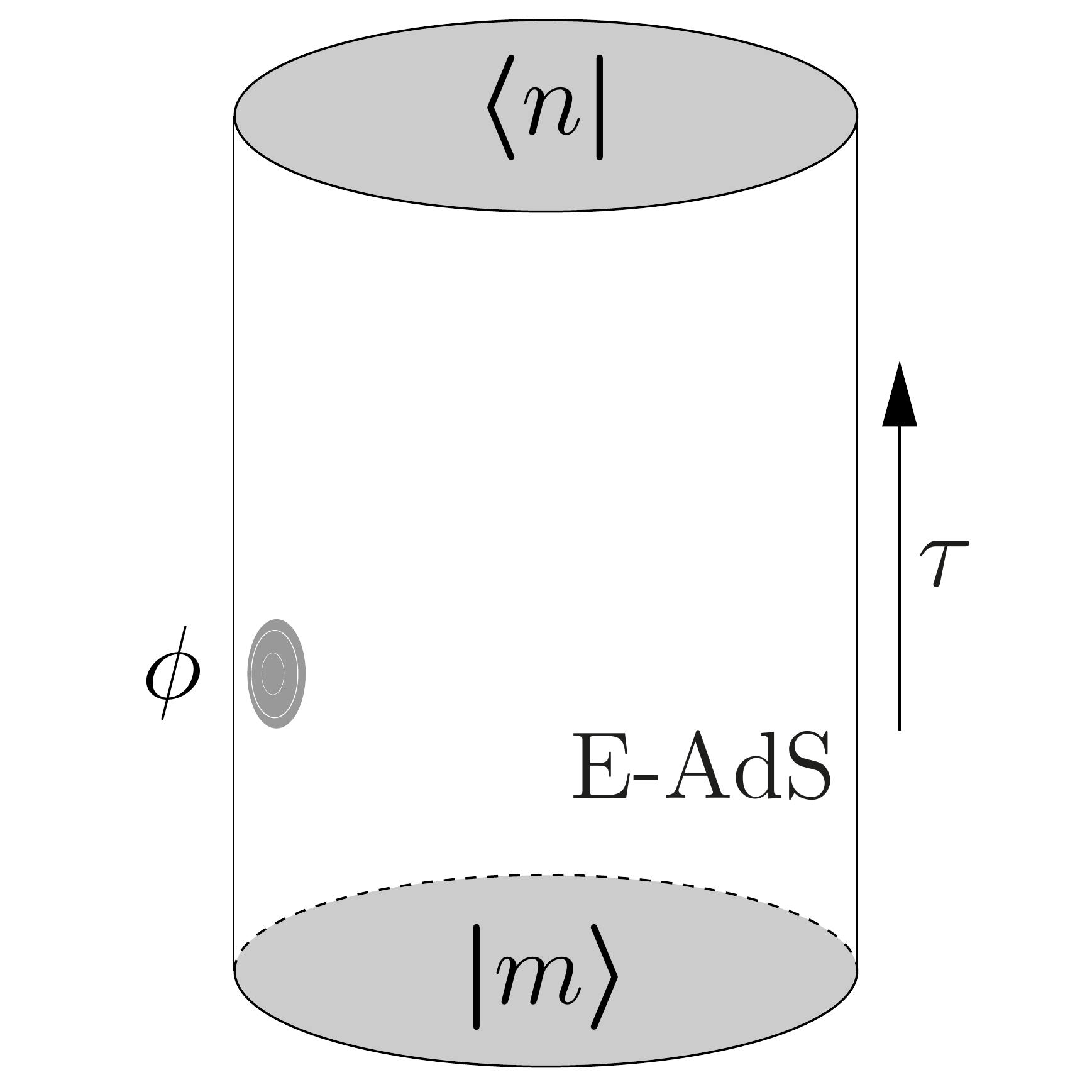}
\caption{}
\end{subfigure}
\begin{subfigure}{0.49\textwidth}\centering
\includegraphics[width=.95\linewidth] {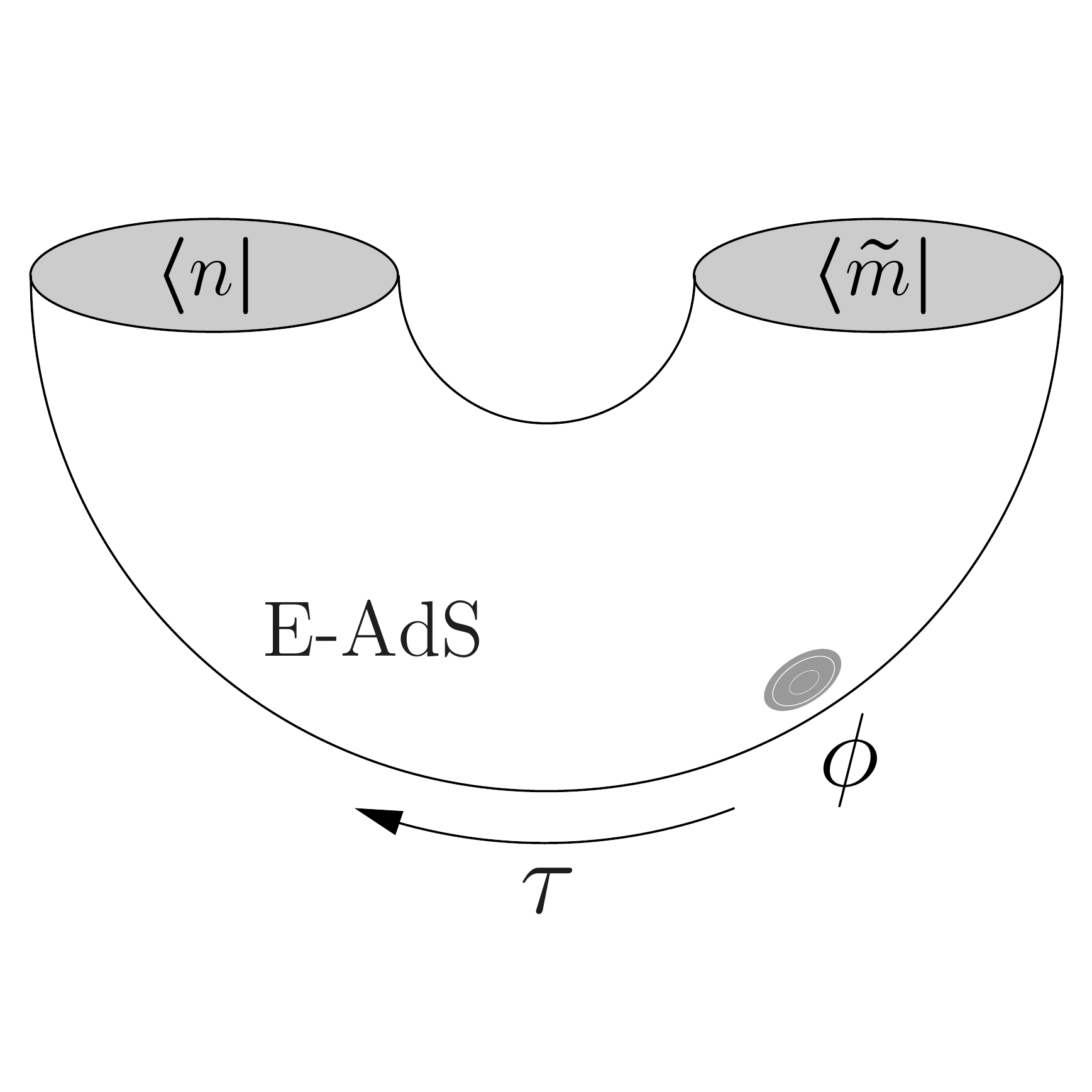}
\caption{}
\end{subfigure}
\caption{(a) A piece of Euclidean evolution depicted as the matrix element $\langle n |\rho_\phi|m\rangle$ of the density matrix $\rho_\phi$. (b) The same object can be also understood as the coefficient $\left( \langle n |\otimes \langle \tilde m |\right)|\Psi_\phi \rangle\!\rangle$ of a ket $|\Psi_\phi \rangle\!\rangle$ defined in the TFD Hilbert space $\cal H \bigotimes\tilde H$.}
\label{Fig:States}
\end{figure}

The connection between the operators in (\ref{Upm}) and a pure state $|\Psi_{\phi} \rangle\!\rangle$ in the TFD framework, arises from the identification of the $U_\phi$ matrix elements in a single Hilbert space, with the components of the state in the doubled Hilbert space \cite{prev}
\be\label{defTFDstate}
\Psi_{\phi}(n, \tilde m )=\left(\langle n|\otimes \langle \tilde m | \right)\;|\Psi_{\phi} \rangle\!\rangle \equiv \langle n |U_{\phi}| m \rangle
\ee
where $| n\rangle, |\tilde m \rangle $ are orthonormal basis of ${\cal H}$ and $\widetilde{{\cal H}}$ respectively. This
expression  is schematically represented in Fig. \ref{Fig:States}: $U_\phi$ is depicted on the left as an evolution operator  on a single Hilbert space, the corresponding TFD-ket  $|\Psi_\phi\rangle\!\rangle$ is illustrated on the right with the two cylinder's ends now representing the d.o.f. of the TFD double intersected at some spacelike surface at a fixed time $t$.

The corresponding {\it bra} state is defined naturally in terms of the adjoint of the operator $U_{\phi}$  
\be  
\Psi^*_{\phi} (n,\tilde{m}) =\Psi_{\phi^*} (n,\tilde{m})\equiv
\langle\!\langle \Psi_\phi |\left(|n\rangle \otimes | \tilde{m} \rangle\right) = \langle m | (U_{\phi})^\dagger |n\rangle  =   \langle m | (U_{\phi^*}) |n\rangle\;,             \ee
where:  $\phi^{*}(\tau)\equiv \phi(-\tau)\; , \;\; \tau \in (-\beta/2\, , 0)$.                                                                

The explicit solution to \eqref{defTFDstate} is
\be
\label{psi-rho-id}
|\Psi_\phi \rangle\!\rangle = (U_\phi\otimes \mathbb I)|1\rangle\!\rangle\;  =U_\phi\,|1\rangle\!\rangle\,,
\ee
where
\be
|1\rangle\!\rangle \equiv\sum_n \,|n\rangle \otimes |\tilde{n} \rangle
\label{unit}\;.
\ee
To show that \eqref{psi-rho-id} satisfies \eqref{defTFDstate} we partially project using   $\langle \tilde{m}| (U_\phi\otimes \mathbb I)|1\rangle\!\rangle = U_\phi\langle \tilde{m}  |1\rangle\!\rangle = U_\phi |m\rangle $. 
In the absence of sources, the state \eqref{psi-rho-id} becomes the TFD vacuum $|\Psi_0\rangle\!\rangle$.
A useful property of \eqref{unit} that will be used in what follows is that 
\be
\label{id2} \text{Tr} \;A =\sum_n \;  \langle n| A  |n\rangle \; \langle \tilde{n}| \tilde{n}\rangle = \,\langle\!\langle  1| A  |1\rangle\!\rangle\;,
\ee
for any operator $A$ acting only on ${\cal H}$. 

\subsubsection{TFD evolution and transition amplitudes}

In order to complete our claim that the TFD construction allows to interpret the SK path integral as an In-Out process, we will show that the transition amplitude from the initial 
state $|\Psi_I \rangle\!\rangle$ to the final state $|\Psi_F \rangle\!\rangle$  is equivalently described by the Schwinger-Keldysh path integral ${\cal C}$, such as in the l.h.s. of eq. (\ref{SvRPath}):
\be\label{inout-Z}
Z_{CFT}({\cal C})=\langle\!\langle \Psi_F | \mathbb{U}(\Delta T=T_F -T_I)\;|\Psi_I\rangle\!\rangle \;.
\ee
We prove this assuming that there is no source in the real time interval, such that the evolution operator of the original system is given by $U_0(t) = e^{-itH}$. The l.h.s. of eq. (\ref{SvRPath}) is, by definition, 
\be\label{ZU}
Z_{CFT}({\cal C}) \equiv \text{Tr} \;  \;U = \text{Tr} \; U_{F}(\beta/2) U_{0}(\Delta T) U_{I}(\beta/2) U_{0}(-\Delta T)
\ee
where the reversed time evolution $U_0(-\Delta T)$ comes from the path ordering.

Therefore,  introducing the TFD double and using \eqref{id2} we  substitute the trace by the expectation value in the unit state, $|n\rangle$ stands for the energy basis. Notice that, according to the second and third rule of eq. (\ref{Tilde-rules})
\be\label{}
 U_0(-\Delta T )|1\rangle\!\rangle = \sum_n e^{i\,\Delta T\, E_n } |n ,\,\tilde{n}\rangle\!\rangle\,=\,\tilde{U_0}(\Delta T) |1\rangle\!\rangle\,\;\,,
 \ee
explicitly $ \tilde{U_0}(\Delta T) = e^{i \Delta T \,\tilde{H} }$. Using also the definition of the initial and final excited states (\ref{psi-rho-id})  for generic sources $\phi_{I,F}$\be     
 U_{I,F}(\beta/2) |1\rangle\!\rangle \equiv \;|\Psi_{I,F}\, \rangle\!\rangle\;,  \ee
 we can express finally (\ref{ZU}) as
\begin{align}
Z_{CFT}({\cal C}_{})
&= \text{Tr} \; U_{F}(\beta/2) \,U_{0}(\Delta T) U_{I}(\beta/2) \, U_{0}(-\Delta T)\nn\\ \nn
&=\langle\!\langle  1| U_{F}(\beta/2) U_{0}(\Delta T) \,U_{I}(\beta/2) U_{0}(-\Delta T) |1\rangle\!\rangle \nn\\
&=\langle\!\langle  1| U_{F}(\beta/2)  U_{0}(\Delta T) \,\tilde{U}_{0}(\Delta T) U_{I}(\beta/2)|1\rangle\!\rangle
\\ \nn
&= \langle\!\langle  \Psi_F  |  \left(U_{0}(\Delta T) \tilde{U}_{0}(\Delta T)\right)|\Psi_I \rangle\!\rangle
\\ \nn
&= \langle\!\langle  \Psi_F  | \mathbb{U}_{0}(\Delta T)|\Psi_I \rangle\!\rangle ;
\end{align}
 where in the second line, we have used the first rule (\ref{Tilde-rules}):
 $[\tilde{U}_{0}(\Delta T)\,,\,U_{I}(\beta/2) ]=0$. In the last line, the evolution operator  of the TFD double (in absence of sources) has been defined as 
\be
\label{Udoble}
\mathbb{U}_0(t) \equiv e^{ - i\; t \;\mathbb{H} } = e^{ -i \, t \,\left( H \otimes \mathbb{1}-\mathbb{1} \otimes H\right)} = U_0(t) \otimes \tilde{U}_0(t)\;,
\ee
where the second term of the extended Hamiltonian is nothing but the operator $\tilde{H}$.
This manifestly shows that the Schwinger-Keldysh CFT partition function expresses a transition amplitude in TFD formalism.

\begin{figure}[t]\centering
\begin{subfigure}{0.49\textwidth}\centering
\includegraphics[width=.9\linewidth] {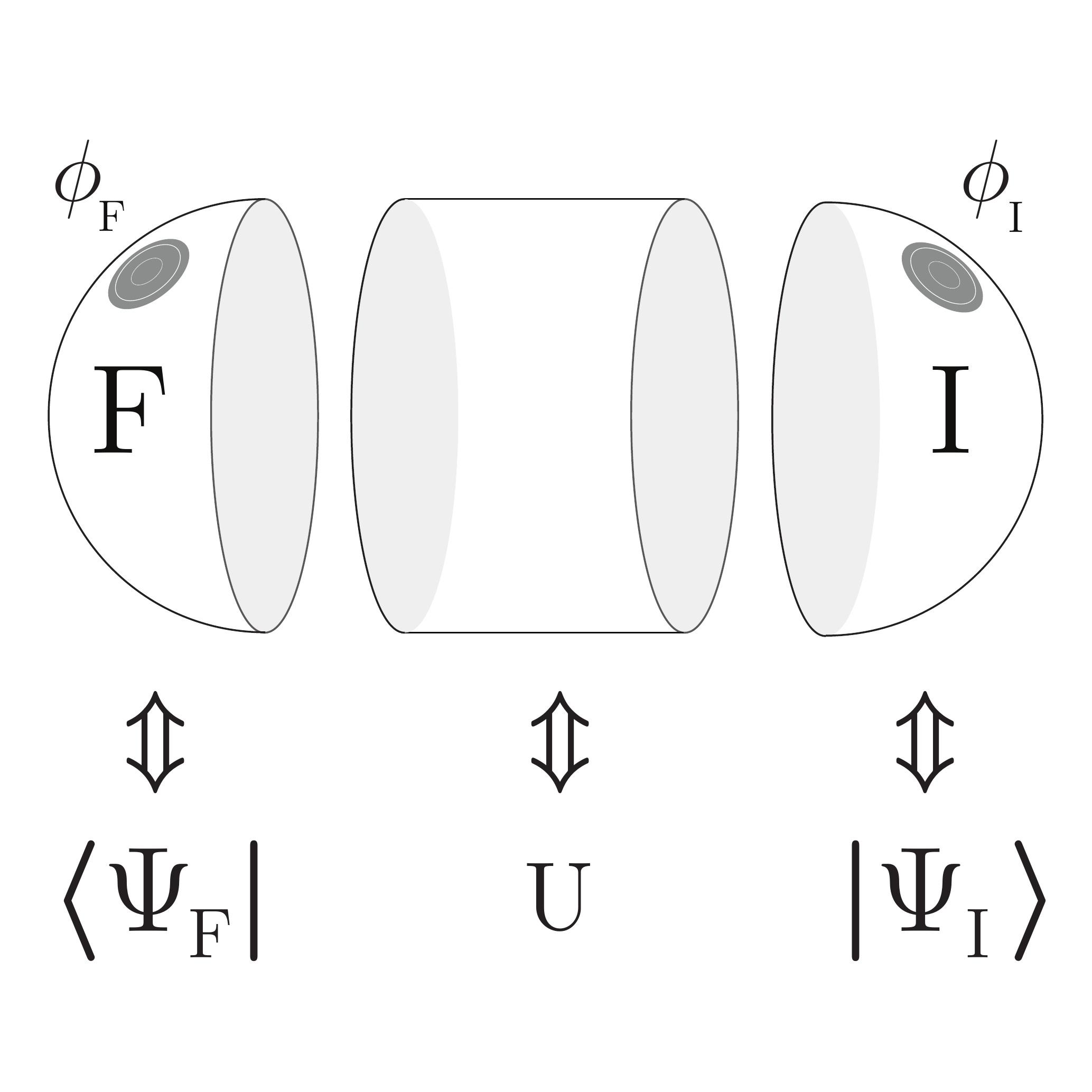}
\caption{}
\end{subfigure}
\begin{subfigure}{0.49\textwidth}\centering
\includegraphics[width=.9\linewidth] {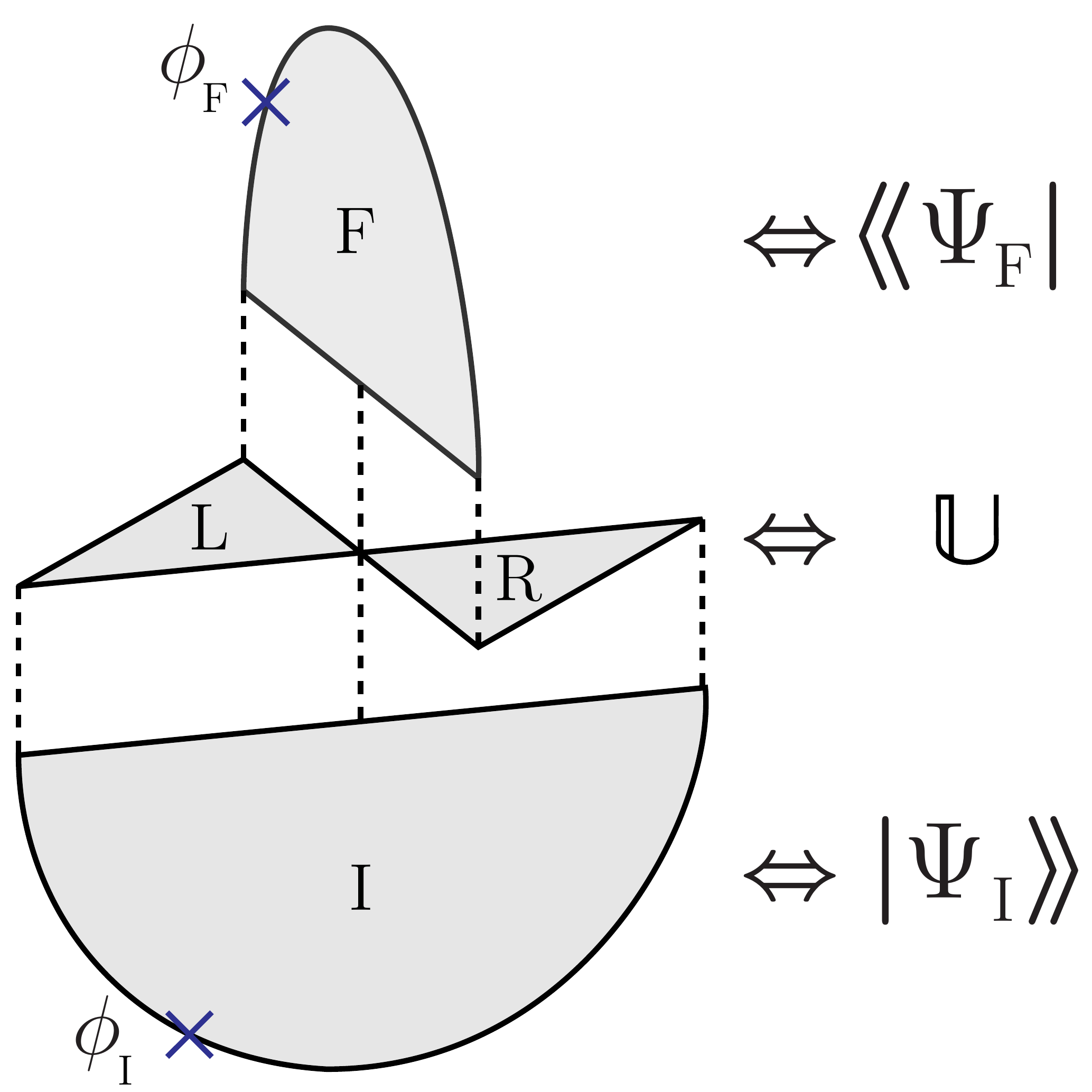}
\caption{}
\end{subfigure}
\caption{The complex-signature manifolds can be split and studied piecewise. Each Euclidean piece corresponds to a state preparation, while the Lorentzian pieces describe the real-time evolution of the system. (a) shows the map at zero temperature, while (b) represents the map at finite temperature.}
\label{Fig:Estado}
\end{figure}

A remarkable conclusion arises from this analysis: the operator (\ref{Udoble}) represents the dual of the Lorentzian part of the geometry shown in Fig \ref{Fig:Estado}(b).  
Notice finally that the vacuum state $|\Psi_0\rangle\!\rangle$
is preserved by the operator (\ref{Udoble}). In other words, the time evolution generated by the Hamiltonian $\mathbb{H}$ is a symmetry for this state\footnote{On the other hand, time evolution given by $\mathbb{H} = \left( H \otimes \mathbb{1}+\mathbb{1} \otimes H\right)$ will not leave the TFD vaccum invariant, see \cite{HM,papa}.}. Nevertheless, we will see later that the gravity dual of the in/out states are coherent, and that an arbitrary coherent state $|\Psi_\phi\, \rangle\!\rangle$ is not invariant but remarkably, the coherence property shall be preserved, namely: 
\be \label{StateEvolution}
|\Psi_{\phi\,'}\,\rangle\!\rangle = \mathbb{U}_0(t)\;|\Psi_{\phi}\, \rangle\!\rangle
\ee
where $\phi\,'=\phi\,'(\phi;t)$.

\subsubsection{Piecewise holographic map }

In this subsection we briefly discuss the structure described above from the bulk perspective, interpreting the different pieces of the geometry of Fig 2(b) as states and propagators.
A well known fact is that there are two dual geometries to Fig. \ref{Fig:Camino}(a), namely those shown in \ref{Fig:Camino}(b) and \ref{Fig:ThermalAdS}(b), which dominate at high and low temperature respectively.  We will focus on the former and refer the reader to App. \ref{App:Thermal} for the computations in the low temperature regime. 

Equation \eqref{inout-Z} from the bulk perspective takes the form 
\begin{equation}\label{TFD-Gravity}
\langle\!\langle \Psi_{F} |\mathbb{U}(\Delta T)|\Psi_{I}\rangle\!\rangle \equiv \int {\cal D}[\Phi]_{\phi({\cal C})} e^{-i S[\Phi]}\,.
\end{equation}
At high temperature, the two Lorentzian pieces of Fig. \ref{Fig:Camino} belong to a single black hole, so that we can split
\begin{equation}\label{path-fig4b}
\int {\cal D}[\Phi]_{\phi({\cal C})} e^{-i S[\Phi]}=
\sum_{\phi_\Sigma(T_I)\,, \phi_\Sigma(T_F)} 
 \left( \int_{\phi_\Sigma(T_F)} {\cal D}[\Phi]_{\phi_F} e^{- S[\Phi]} \right)
 \left( \int_{\phi_\Sigma( T_I)}^{\phi_\Sigma(T_F)} {\cal D}[\Phi]_{\phi_L , \phi_R} e^{-i S[\Phi]} \right)
 \left( \int^{\phi_\Sigma( T_I)} {\cal D}[\Phi]_{\phi_I} e^{- S[\Phi]} \right)
\end{equation}
where the subindex in ${\cal D}[\Phi]_{\phi}$ stand for asymptotic sources. We have also denoted by $\phi_\Sigma( T_I) := \{ \Phi(x) \;, \; x\in \Sigma( T_I) \}$ the (complete) $\hat \Phi$ configuration basis for a smooth hypersurface $\Sigma$ at $ T_I$ intersecting the asymptotic boundary in two disconnected spheres $\partial \Sigma = S^{d-1} + S^{d-1}$ \cite{eternal}.  The topology of $\Sigma$,  homologous to the (Euclidean) asymptotic boundary \cite{Aitor-Simplectic,Maloney:2015ina}, is that of the Einstein-Rosen bridge $\Sigma=S^{d-1} \times I$, with $I$ the interval for the holographic coordinate \footnote{ 
At low temperatures, there is no bulk connection between the boundaries and the path integral is performed over two separate AdS geometries. In that set-up, see Fig. \ref{Fig:ThermalAdS}, the dofs naturally split.}.
From the CFT viewpoint there is a similar decomposition
\begin{equation}
\langle\!\langle \Psi_F | \mathbb{U}(\Delta T) \; |\Psi_I\rangle\!\rangle = \sum_{\phi_\Sigma(T_I),\phi_\Sigma(T_F)} \langle\!\langle \Psi_F |\phi_\Sigma(T_F)\rangle\!\rangle\langle\!\langle \phi_\Sigma(T_F)| \mathbb{U} (\Delta t) |\phi_\Sigma(T_I)\rangle\!\rangle \; \langle\!\langle \phi_\Sigma( T_I) |\Psi_I\rangle\!\rangle\;.
\end{equation}
Comparing the equations above it is natural to identify bulk path integrals with CFT expressions as: 
\be\label{lorentzianpath}
\langle\!\langle \phi_\Sigma(T_F) | \mathbb{U}_{}(\Delta T)
|\phi_\Sigma( T_I)\rangle\!\rangle \equiv  \int_{\phi_\Sigma(T_I)}^{\phi_\Sigma(T_F)} {\cal D}[\Phi]_{\phi_L , \phi_R} e^{-i S[\Phi]} \;,\ee
while the states are prepared by
\begin{equation}\label{bulk-states}
\langle\!\langle \phi_\Sigma(T_I)|\Psi_I\rangle\!\rangle \equiv 
 \int^{\phi_\Sigma(T_I)} {\cal D}[\Phi]_{\phi} e^{- S[\Phi]} 
\qquad\qquad 
\langle\!\langle \Psi_F |\phi_\Sigma(T_F)\rangle\!\rangle \equiv \int_{\phi_\Sigma(T_F)} {\cal D}[\Phi]_{\phi} e^{- S[\Phi]}  \;.
\end{equation}
This map is illustrated in Figs. \ref{Fig:Estado} in complete analogy with the zero-temperature scenario \cite{us}.

In our previous work \cite{prev} we gave a dual description of the Lorentzian operator $\mathbb{U}(\Delta T)$ in terms of bulk boost-like evolution of two exterior wedges of a single black hole. 
Regarding the states, it is well known that in the absence of sources they correspond to TFD vacuum states \cite{eternal}. Turning on sources will create excitations over these backgrounds which we will study holographically in the forthcoming sections. 

We would like to conclude this section by writing the resulting holographic prescription in the semi-classical limit of gravity. From \eqref{TFD-Gravity} and \eqref{path-fig4b}, we have
\begin{equation}\label{TFD-Gravity-largeN}
\langle\!\langle \Psi_{F} |e^{i\int_{T_I}^{T_F} \left({\cal O}_L \phi_L - {\cal O}_R\phi_R\right)}|\Psi_{I}\rangle\!\rangle \approx e^{-i S^0[\phi({\cal C})]} 
\end{equation}
where, for consistency with previous literature \cite{SvRC} we have expressed the evolution operator $\mathbb{U}(\Delta T)$ in the lhs in the interaction picture.  This is the relation we will work with in the rest of the paper.

\section{Excited states from the bulk perspective}
\label{Sec:Bulk}

The aim of this section is to compute matrix elements of local operators ${\cal O}$, with mass dimension $\Delta$, between excited states obtained from non-zero sources in the Euclidean sections, and inner products between excited states in the geometry shown in Fig. \ref{Fig:Camino}(b). This is achieved in a standard semi-classical approach by solving the bulk EOMs for general sources and evaluating the on-shell action. The results represents the CFT behavior for high temperature. The low temperature CFT behavior is obtained from Thermal-AdS geometry. We relegate its study to App. \ref{App:Thermal}.

\subsection{Bulk geometry and gluing conditions}

\begin{figure}[ht]\centering
\begin{subfigure}{0.49\textwidth}\centering
\includegraphics[width=.9\linewidth] {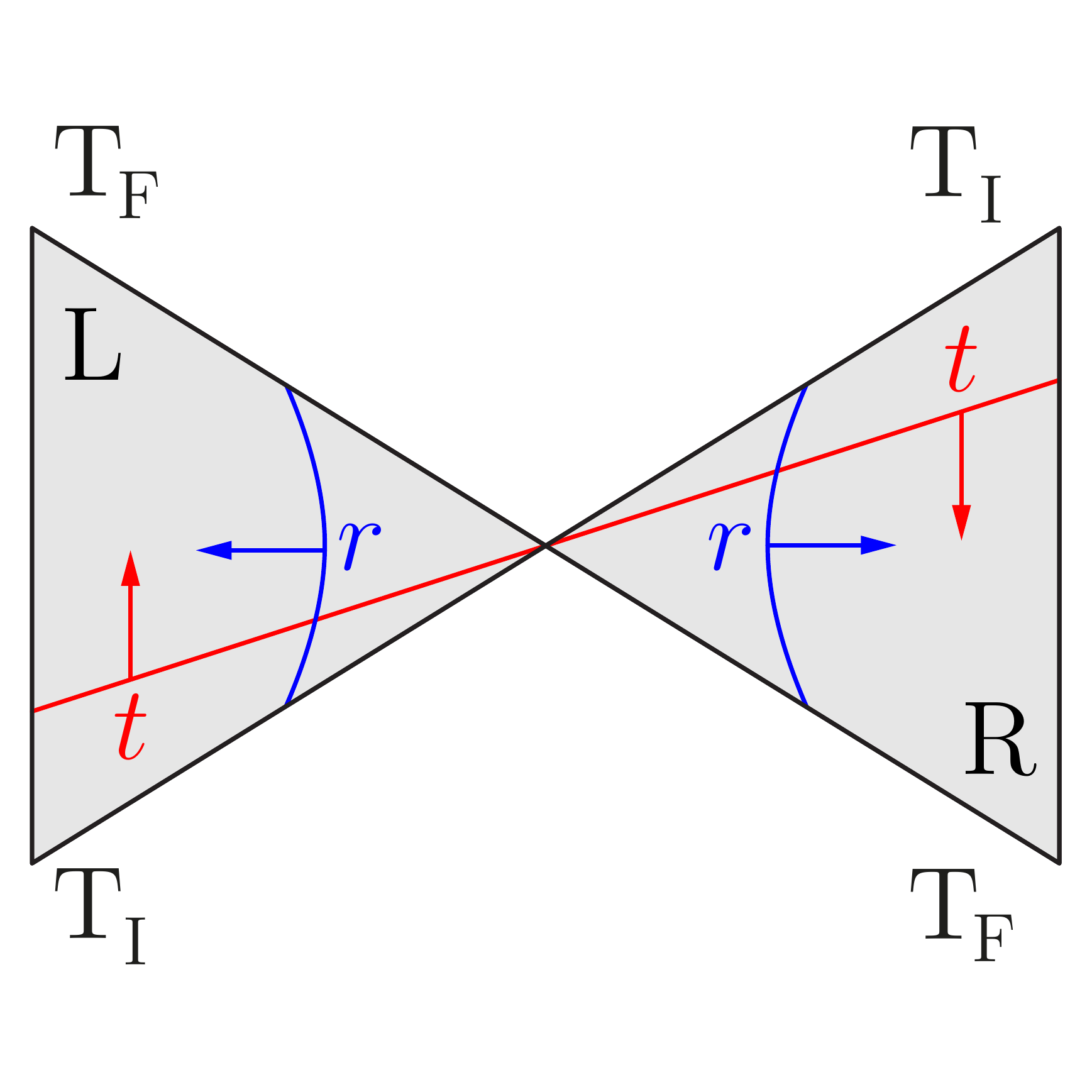}
\caption{ }
\end{subfigure}
\begin{subfigure}{0.49\textwidth}\centering
\includegraphics[width=.9\linewidth] {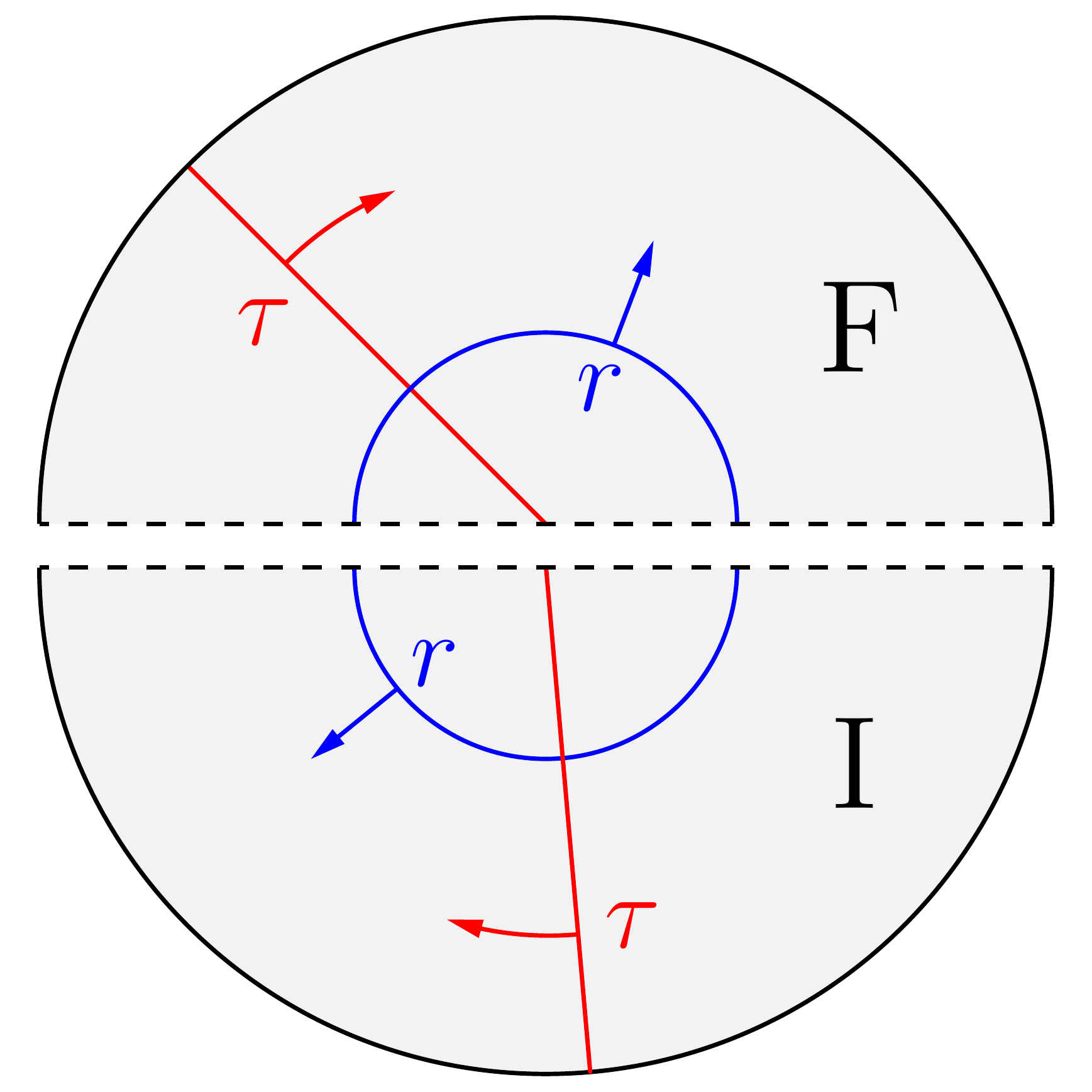}
\caption{}
\end{subfigure}
\caption{(a) Static patches of the AdS-BH with constant $t,r$ surfaces depicted. Time runs upward in the left wedge (L) and downward in the right wedge (R). The angular variable $\varphi$ in \eqref{metric-l2} has been suppressed. 
(b) Euclidean AdS-BH: time becomes an angular variable $\tau\sim\tau+2\pi$. The two pieces are identical and their temporal extension is $\beta/2$. }
\label{Fig:Guille}
\end{figure}

The geometry is built from a static Lorentzian AdS-BH exterior and an Euclidean BH manifold halved in two pieces, see fig.\ref{Fig:Guille}. The two Euclidean pieces are glued along constant $t$-hypersurfaces located at $t=T_I$ and $t=T_F$ shown as red lines in fig. \ref{Fig:Guille}(a). We work with a 3d bulk, nevertheless our ideas extend straightforwardly to higher dimensions. The Lorentzian and Euclidean metrics are \cite{BTZ}
\begin{equation}\label{metric-l2}
d s^2=-(r^2-1)d t^2+\frac{d r^2}{(r^2-1)}+r^2 d\varphi^2 \qquad\qquad d s^2=(r^2-1)d\tau^2+\frac{d r^2}{(r^2-1)}+r^2 d\varphi^2
\end{equation}
with $\tau\sim \tau +2\pi$. We have mapped the BH temperature parameter into the angular periodicity $\varphi\sim\varphi+2\pi r_S$.

Without loss of generality we take the Lorentzian time extension $t\in [T_I,T_F]=[-T/2,T/2]$ (cf. complex path in Fig. \ref{Fig:Camino}(a)). Notice that the path ordering and the time coordinate run in opposite directions in the R-wedge, consistent with the TFD interpretation \cite{Israel}. For the Euclidean regions we choose $\tau\in[-\pi,0]$ in region I and $\tau\in[0,\pi]$ in region F.  

The gluing conditions between the regions follow from a saddle point approximation of eq. \eqref{SvRPath}, which demand  ${\cal C}^1$ continuity of the fields across the gluing regions. The conditions are
\begin{eqnarray}\label{bc}
\Phi_L=\Phi_I\,,  & \quad
-i\partial_t\Phi_L=\partial_\tau\Phi_I\,, 
& \qquad\text{on $t=T_I$, $\tau=0$}\nn \\
\Phi_L=\Phi_F\,, & \quad
-i\partial_t\Phi_L=\partial_\tau\Phi_F\,, 
& \qquad\text{on $t=T_F$, $\tau=0$} \nn\\
\Phi_R=\Phi_I\,,& \quad
-i\partial_t\Phi_R=\partial_\tau\Phi_I\,, 
&\qquad\text{on $t=T_I$, $\tau=-\pi$} \nn\\
\Phi_R=\Phi_F\,,& \quad
-i\partial_t\Phi_R=\partial_\tau\Phi_F\,, 
& \qquad\text{on $t=T_F$, $\tau=\pi$}\;.
\end{eqnarray}
A detailed discussion on the gluing of the geometry pieces themselves was presented in \cite{prev}.

\subsection{Scalar field solution}

We will now find the classical solution for a free real massive scalar field   subject to arbitrary boundary conditions on the asymptotic region of the manifold depicted in Fig. \ref{Fig:Camino}(b). The resulting solution will show up a non-trivial feature: only the choice of identical $\beta/2$ Euclidean pieces guarantee the analyticity of the solution through the wormhole in the complex $r$-plane when general sources are turned on.
This was stated in \cite{prev}, we demonstrate it below and further discuss its consequences in Sec. \ref{Sec:Analitycity}.

The linearity of the problem allows to build the general solution out of a linear combinations of solutions with non-zero sources on a single region. Thus, we begin by building the solution for a Lorentzian source. We then build the solution with non-zero Euclidean sources highlighting the relevant differences with the Lorentzian case. We will also show that our solution can be easily related to the standard Unruh-like basis $\phi_{\pm\pm}$ discussed in \cite{SvRL,herzog}.

\subsubsection{Lorentzian Sources}
 
\begin{figure}[t]\centering
\begin{subfigure}{0.49\textwidth}\centering
\includegraphics[width=.9\linewidth] {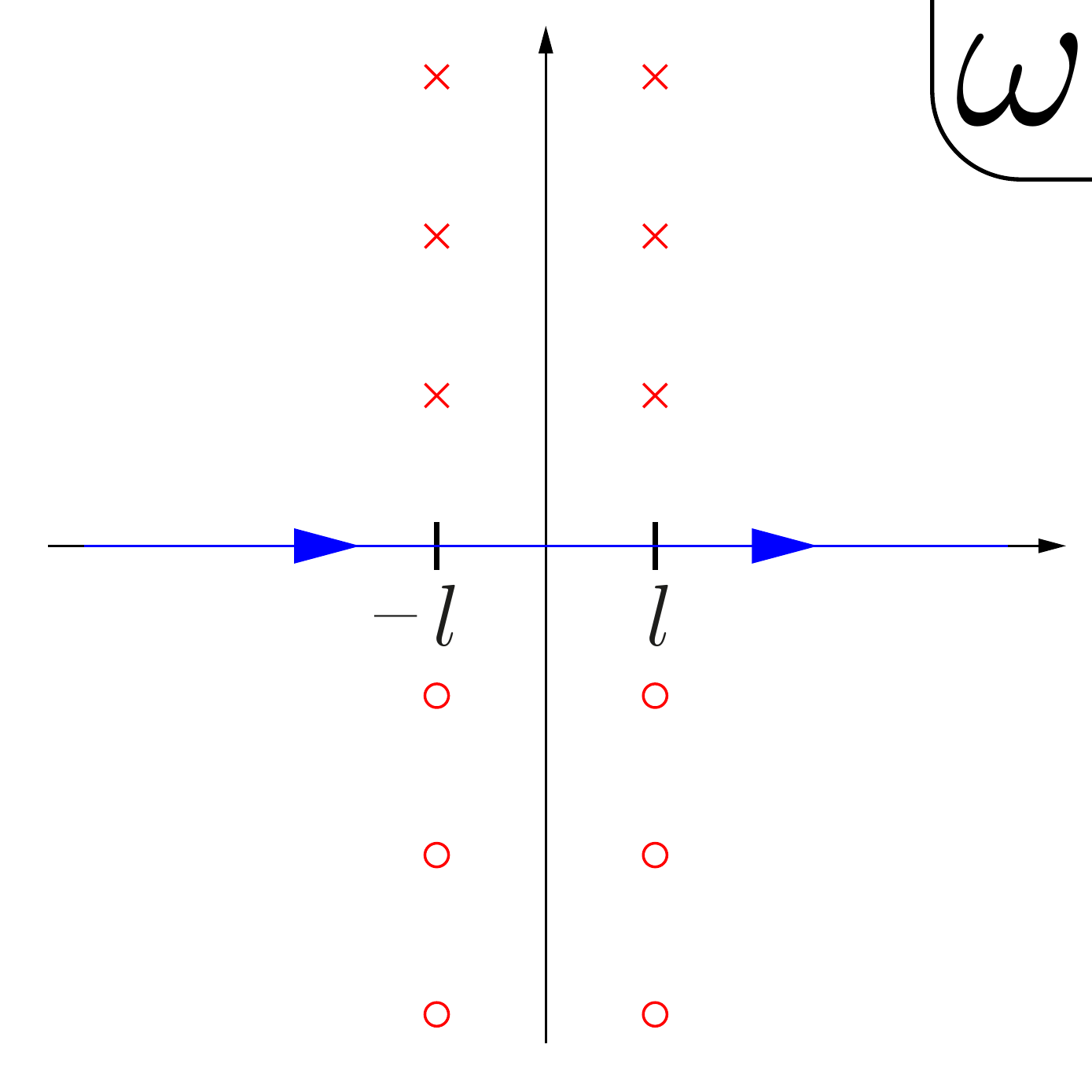}
\caption{}
\end{subfigure}
\begin{subfigure}{0.49\textwidth}\centering
\includegraphics[width=.9\linewidth] {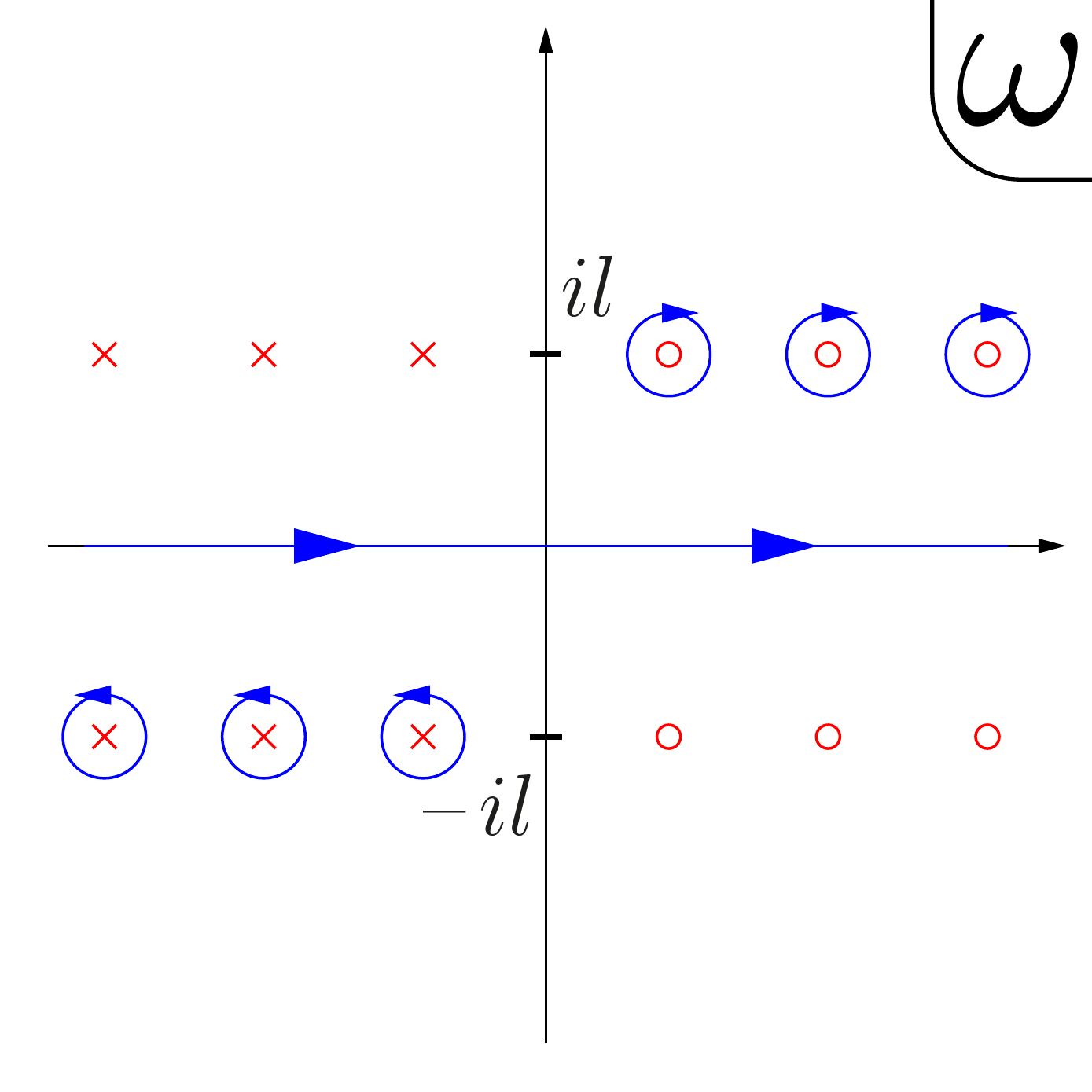}
\caption{}
\end{subfigure}
\caption{(a) Crosses show the location of the poles of $f(\omega,l,r)$ while circles those of $f(-\omega,l,r)$.  The blue line denotes the $\omega$-integration contour in \eqref{Lsol}. 
(b) Location of the poles of $f(\pm i\omega,l,r)$ for the treatment of Euclidean sources. The contour runs along the real axis and the QN$(r,\tau,\varphi)$ contribution in \eqref{qqn} arise from the encircled poles. Crosses denote poles of $f(-i\omega,l r)$ while circles denotes those of $f(i\omega,l r)$. Rotating the contour in fig.(b) to the imaginary axis counterclockwise reduces \eqref{qqn} to \eqref{Isol2c}, recovering fig.(a).}
\label{Fig:Polos}
\end{figure}

The action and equations of motion for the L-region are
\begin{equation}\label{scalarEOM}
S[\Phi]=-\frac 12 \int dt dr d\varphi \sqrt{|g|}\left(\partial_\mu\Phi\partial^\mu\Phi+m^2\Phi^2\right)\;,\qquad\qquad\left(\square-m^2\right)\Phi=0 \;,\qquad\qquad m^2=\Delta(\Delta-2)
\end{equation}
in the metric \eqref{metric-l2}. Expanding in plane waves as $\Phi = e^{-i \omega t + i l \varphi} f(\omega,l,r)$, where $l\in\mathbb{Z}/r_S$ one obtains 
\begin{equation}
\frac{1}{r}\partial_r\left[r(r^2-1)\partial_r\right]f(\omega,l,r)=\left[m^2+\frac{l^2}{r^2}-\frac{\omega^2}{r^2-1}\right]f(\omega,l,r)\,.
\end{equation}
The two linearly independent solutions are $f(\pm\omega,l,r)$, where 
\begin{equation}\label{f}
f(\omega,l,r)\equiv {\cal C}_{\omega l\Delta} \; r^{-\Delta } \left(1-\frac{1}{r^2}\right)^{i\frac{\omega }{2}} \, _2F_1\left(\frac{\Delta }{2}+\frac{1}{2} i (\omega -l),\frac{\Delta }{2}+\frac{1}{2} i
   (\omega+l );i \omega +1;1-\frac{1}{r^2}\right)\,,
\end{equation}
\begin{equation}\nn
{\cal C}_{\omega l\Delta}\equiv \frac{\Gamma \left(\frac{\Delta }{2}+\frac{1}{2} i (\omega -l)\right) \Gamma \left(\frac{\Delta }{2}+\frac{1}{2} i (\omega +l)\right)}{\Gamma (\Delta -1) \Gamma (i \omega +1)}\,.
\end{equation}
The normalization factor is set so that\footnote{The $\ln(r^2)$ term in \eqref{fexpansion} appears only for $\Delta\in\mathbb{N}$ and becomes relevant in KK compactifications. This will not be relevant for our discussion. We refer the interested reader to \cite{DHokerFriedman} and appendices in \cite{us2}.}
\begin{equation}\label{fexpansion}
f(\omega,l,r)\approx r^{\Delta-2}+\dots+ \alpha_{\omega l\Delta} r^{-\Delta}\left[\ln(r^2) +\beta_{\omega l\Delta} + \dots\right]\,,\qquad r\to\infty
\end{equation}
\begin{equation}\label{alpha}
\alpha_{\omega l\Delta}\equiv (-1)^{\Delta -1} \frac{ \left(\frac{2-\Delta }{2}+\frac{i}{2}  (\omega - l)\right)_{\Delta -1} \left(\frac{2-\Delta }{2}+\frac{i}{2}  (\omega + l)\right)_{\Delta -1}}{(\Delta -2)! (\Delta -1)!}\,,
\end{equation}
\begin{equation}\label{beta}
\beta_{\omega l\Delta}\equiv-\psi \left(\frac{\Delta }{2}+\frac{i}{2}  (\omega -l)\right)-\psi \left(\frac{\Delta }{2}+\frac{i}{2}  (\omega +l)\right)\,,
\end{equation}
where $(x)_y$ stands for Pochhammer symbols and $\psi(x)$ for Digamma functions. The analytic structure of the solutions show simple poles at $\omega=\pm l +i (2n+\Delta)$, with $n\in\mathbb{N}$ arising from ${\cal C}_{\omega l\Delta}$, see fig. \ref{Fig:Polos}(a). 

In the BH context one thus finds two linearly independent regular NN solutions:  $ e^{-i \omega t + il\varphi } f(\pm\omega,l,r)$. They correspond to purely outgoing and infalling modes at the horizon, respectively. The general solution on the L-region is then,
\begin{equation}\label{Lsol}
\Phi_L(r,t,\varphi)=\frac{1}{4\pi^2 r_S} \sum_{l} \int d\omega 
\; e^{-i\omega t+i l \varphi}\; \bar \phi_L(\omega,l)\left[L^+_{\omega l} f(\omega,l,r) + L^-_{\omega l} f(-\omega,l,r)\right]\;,
\end{equation}
where $\omega\in\mathbb R$, the sum is implicit over $l\in\mathbb{Z}/r_S$, $\bar \phi_L(\omega,l)$ is the Fourier transform of the source $\phi_L(t,\varphi)$ and 
\be 
L^+_{\omega l} + L^-_{\omega l}=1\,.
\label{Lpm}
\ee
This last condition is required to meet the asymptotic boundary condition $\Phi_L(r,t,\varphi)\to r^{\Delta-2}\phi(t,\varphi)+\dots $ as $r\to\infty$ . Introducing $L^\pm_{\omega l}$ becomes handy for gluing the complete solution. 
To gain some more physical intuition, we notice that the quotient $L^+_{\omega l}/L^-_{\omega l}$ gives the relative weight of outgoing and infalling modes through the horizon in the NN solution. 

The N modes are built from the combination $e^{-i \omega t +il\varphi } \left[f(\omega,l,r)-f(-\omega,l,r)\right]$ as can be seen from \eqref{fexpansion}\footnote{See \cite{kenmoku} for a discussion of normalizable modes in BTZ geometry}. One can then think of $L^+_{\omega l} - L^-_{\omega l}$ as defining the N modes content of the solution on the L-region. The solutions for $\Phi$ on regions R-, I- and F-regions are expanded in N modes as
\begin{align}
\Phi_R(r,t,\varphi)&= \frac{1}{4\pi^2 r_S} \sum_{l} \int d\omega \; e^{-i\omega t+i l \varphi} R_{\omega l} \left[ f(\omega,l,r) - f(-\omega,l,r)\right] \label{Rsol}\;,\\
\Phi_{I}(r,\tau,\varphi)&= \frac{1}{4\pi^2 r_S} \sum_{l} \int d\omega  \; e^{-\omega \tau+i l \varphi} \; I_{\omega l} \left[ f(\omega,l,r) - f(-\omega,l,r)\right]  \label{Isol}\;,\\
\Phi_F(r,\tau,\varphi)&= \frac{1}{4\pi^2 r_S} \sum_{l} \int d\omega \; e^{-\omega \tau+i l \varphi} \; F_{\omega l} \left[ f(\omega,l,r) - f(-\omega,l,r)\right] \label{IIsol}\;.
\end{align}
The $\omega$-integrals in the last two expressions look divergent at its endpoints, but it turns out that the final coefficients $I_{\omega l}$ and $F_{\omega l}$ keep them regular.

The gluing is now performed profiting from the analytic structure of \eqref{f} in the complex $\omega$-plane. As illustrated in fig.\ref{Fig:Polos}(a) by Residues Theorem one has
\begin{equation} \label{intf=0}
\int d\omega  \;e^{-i\omega \Delta t}  f(\omega,l,r)
=0\,,\qquad\qquad\Delta t>0\,.
\end{equation}
Consider the gluing between regions L and F at $t\sim T/2$: the source  $\phi_L(t',\varphi')$ has support to the past of the gluing surface thus making $\Delta t=t-t'>0$ in \eqref{Lsol}. Inserting \eqref{Lpm} and using \eqref{intf=0} one finds
\begin{align}\label{tLT}
\Phi_L(r,t,\varphi)
&=\frac{1}{4\pi^2 r_S} \sum_{l} \int d\omega \; e^{-i\omega t +i l \varphi}  \left(- \bar \phi_L(\omega,l) \;L^-_{\omega l}\right) \left[f(\omega,l,r)-f(-\omega,l,r)\right]\;,~~~~ {t\sim T/2}
\end{align}
with $\bar\phi_L$ the Fourier transform of the source.
Analogously, for the gluing of L and I at  $t\sim -T/2$ one finds
\begin{align}\label{tL0}
\Phi_L(r,t,\varphi)
&=\frac{1}{4\pi^2 r_S} \sum_{l} \int d\omega \; e^{-i\omega t +i l \varphi}  \left( \bar \phi_L(\omega,l) \;L^+_{\omega l}\right) \left[f(\omega,l,r)-f(-\omega,l,r)\right]\;,~~~~{t\sim -T/2}
\end{align}
It is worth mentioning two important features: (i) \eqref{tLT} and \eqref{tL0} show that the field consists solely of N-modes at the gluing surfaces, and (ii) the quotient $L^+_{\omega l}/L^-_{\omega l}$ determines the causal properties of the solution, e.g. the case $L^+_{\omega l}=0$ and $L^-_{\omega l}=1$ gives the retarded solution discussed in \cite{son,herzog}. 

Using \eqref{Rsol}-\eqref{IIsol}, \eqref{tLT} and \eqref{tL0} the gluing conditions \eqref{bc} give
\begin{align}\label{CoeffSol}
- L^-_{\omega l} \; \bar\phi_L(\omega,l) \; e^{-i \omega T/2}&=F_{\omega l}\;,
&~& 
F_{\omega l}\;e^{-\pi\omega}= \; R_{\omega l}\; e^{-i \omega T/2}\;, 
&~&
R_{\omega l}e^{i \omega T/2}  =I_{\omega l}\;e^{\pi\omega}\;,
&~& 
I_{\omega l}=L^+_{\omega l}e^{i \omega T/2}  \; \bar\phi_L(\omega,l) 
\end{align}
yielding via \eqref{Lpm}
\begin{equation}\label{AandB}
L^+_{\omega l}=\frac{-1}{e^{2\pi\omega}-1} \,, \qquad\qquad L^-_{\omega l}=\frac{e^{2\pi\omega}}{e^{2\pi\omega}-1}\;.
\end{equation}
This expressions were used in \cite{prev} to study correlators in the geometry. From the above equations one can also extract the relations between the coefficients in the L and R regions,
\begin{equation}\label{coeffsL}
R_{\omega l}=  \bar\phi_L(\omega,l)\; L^+_{\omega l}\; e^{\omega \pi }=-\;\bar\phi_L(\omega,l)\; L^-_{\omega l}\; e^{-\omega \pi }\,.
\end{equation}
which we will analyze more in depth after we build the solution with non-zero Euclidean sources.

The study of excited states will require the expressions for the bulk field in the Euclidean sections,
\begin{align}
\Phi_{I}(r,\tau,\varphi)&= \frac{1}{4\pi^2 r_S} \sum_{l} \int d\omega \; e^{-\omega \tau+i l \varphi} \; \left(-\bar\phi_L(\omega,l)\; e^{+i\omega T/2} \frac{1}{e^{2\pi\omega}-1}\right) \left[ f(\omega,l,r) - f(-\omega,l,r)\right] \nn\;,\quad \tau\in(-\pi,0  )\\
\Phi_F(r,\tau,\varphi)&= \frac{1}{4\pi^2 r_S} \sum_{l} \int d\omega  \; e^{-\omega \tau+i l \varphi} \; \left(-\bar\phi_L(\omega,l)\; e^{-i\omega T/2} \frac{e^{2\pi\omega}}{e^{2\pi\omega}-1}\right) \left[ f(\omega,l,r) - f(-\omega,l,r)\right]  \nn\;,\quad \tau\in(0,\pi)
\end{align}
As anticipated, the Boltzmann factors in the resulting coefficients adequately regulate the $\omega$ integrals, thus validating our procedure.

\subsubsection{Euclidean sources }

In this section we consider a non-zero source on the asymptotic boundary of region F. The bulk field consists of N-modes in the R, I and L regions. Then, in addition to \eqref{Rsol} and \eqref{IIsol} we now have
\begin{align}
\Phi_L(r,t,\varphi)&= \frac{1}{4\pi^2 r_S} \sum_{l} \int d\omega  e^{-i\omega t+i l \varphi} L_{\omega l} \left[ f(\omega,l,r) - f(-\omega,l,r)\right] \label{Lsol2}\;.
\end{align}
The solution in region F is written as
\begin{align}
\Phi_F(r,\tau,\varphi)&= \frac{i}{4\pi^2 r_S} \sum_{l} \int  d\omega  \; e^{-\omega \tau + i l \varphi} \bar \phi_{F}(-i\omega,l) \left[F^+_{\omega l} f(\omega,l,r) + F^-_{\omega l} f(-\omega,l,r)\right],  
\label{Isol2c}
\end{align} 
with $F^+_{\omega l}+F^-_{\omega l} = 1$. This ansatz appropriately meets the asymptotic boundary condition. To see this, deform the $\omega$-contour clockwise 
encircling some poles in the procedure. One is left with
\begin{align}
\Phi_F(r,\tau,\varphi)
&= \frac{1}{4\pi^2 r_S} \sum_{l} \int d\omega\;  e^{i \omega \tau+i l \varphi} \bar\phi_{F}(-\omega,l) \left[F^+_{-i\omega l} f(-i\omega,l,r) + F^-_{-i\omega l} f(i\omega,l,r)\right] + \text{QN}(r,\tau,\varphi)\,, \label{qqn}
\end{align}
where QN collects the residues from the poles depicted in fig.\ref{Fig:Polos}(b) and correspond to quasi-normal modes \cite{solo}. These contributions decay asymptotically as $r^{-\Delta}$, so the terms in brackets guarantee the bc provided $F^+_{-i\omega l}  + F^-_{-i\omega l}=1 $. 

The parametrization \eqref{Isol2c} can be easily glued to solutions \eqref{Lsol2} and \eqref{Rsol}.
Proceeding as above, the complete gluing leads to
\begin{equation}
F^+_{\omega l}=\frac{e^{2\pi\omega}}{e^{2\pi\omega}-1}\;,
\qquad
F^-_{\omega l}=\frac{-1}{e^{2\pi\omega}-1}\;,
\qquad
e^{i\omega T/2} L_{\omega l} =I_{\omega l}= i \;\bar \phi_{F}(-i \omega,l) \frac{e^{i\omega T}}{e^{2\pi\omega}-1}
\ee 
and 
\be R_{\omega l}=e^{\omega \pi}L_{\omega l}\; \label{coeffsI}\;.
\end{equation}
Again, Boltzmann factors in the coefficients make the the $\omega$-integral in \eqref{Isol2c} convergent. Moreover, each mode in \eqref{Isol2c} is regular at the horizon. 

The general solution with non-zero sources in all regions can be straightforwardly built from the cases studied above by superposition. Before studying the on shell action, we make some comments on the analytic properties of the solution. 

\subsubsection{Analyticity through the wormhole}\label{Sec:Analitycity}

So far we have built the solutions by gluing Rindler-like modes across spacelike regions of alternating Euclidean and Lorentzian signatures, see \eqref{bc}. As an outcome, the procedure has given a precise connection between the coefficients in the L- and R-regions, this is, through the wormhole.  

We would like to stress that the $\beta/2$ lengths of the Euclidean regions turn out to relate the Lorentzian pieces coefficients L and R by $e^{\pm\omega \pi}$ factors, see \eqref{coeffsL} and \eqref{coeffsI}. These are precisely the standard relations that make the combination of L and R Rindler modes become Unruh-like, i.e. global analytic \cite{Unruh,KeskiVakkuri,Israel}.  Had we chosen $\sigma\neq\beta/2$ this would no longer hold. We follow the notation in \cite{herzog,SvRL} where the Unruh modes were denoted as $\phi_{\pm\pm}$. The first $\pm$ denote pure infalling/outgoing modes in the L quadrant while the second $\pm$ refers to whether the mode is analytic in the lower/upper $U$-complex plane. The `box' character of AdS makes each of these four modes $\phi_{\pm\pm}$ divergent at the asymptotic boundary. Then, the N-modes in Lorentzian signature arise from adequate linear combinations\footnote{As this work was near completion, we became aware of \cite{Glorioso:2018mmw}, which also stresses the analytical properties of the field on the radial holographic coordinate.}.

From \eqref{coeffsI} we see that turning on sources on F awakens $\phi_{\pm,+}$ modes. It is important to stress that the relation  \eqref{coeffsI} follows solely from the gluing of the three sourceless pieces I, L, R. Conversely, it is easy to see that I-sources turn on $\phi_{\pm,-}$ modes. 
Crucially, the resulting combination of (Lorentzian) Rindler modes awakened across the Einstein-Rosen bridge become global N modes. Furthermore the combination is the one related with positive energy Unruh-like particles. We thus recover the intuition that Euclidean sources explore excited Hilbert space states and provide a concrete description of them. We will deepen this analysis in sections to come. For completeness, from \eqref{coeffsL} we quote that L-sources turn on $\phi_{+,+}$ and $\phi_{-,-}$ modes while R-sources excite $\phi_{+,-}$ and $\phi_{-,+}$. However, these (NN) modes are not associated with particle excitations.

The bottom line is: equal $\beta/2$ Euclidean pieces imply that the Lorentzian regions of the geometry should be understood as the L and R wedges of a single maximally extended BH. Within this framework, the study of a finite temperature CFT requires only holographic (boundary) data. The observation that the maximally extended BH is related to identical $\beta/2$ Euclidean segments in the CFT computations was noticed in \cite{herzog}.

\subsection{Results from bulk analysis}\label{Correlators}

From the results in our previous sections one can obtain the full on-shell action 
\begin{align}
i S^0\left[\phi\right]&=-  \frac i2 \int_{\partial}  \sqrt{\gamma} \;\Phi\; n^\mu \partial_\mu \Phi \label{Onchell}=-  \frac{i}{2}  r^{\Delta} \int_{\partial} \;\phi({\cal C}) \left[r\partial_r\Phi\right]_{r\to\infty}\,.
\end{align}
In the following we will present the results for the high and low temperature limit. The computations for the Thermal geometry (low temperature) are straightforward and for completeness we have relegated them to App. \ref{App:Thermal}.

We start computing the inner product between the excited states. In the high temperature regime taking $\Delta T\to0$ in \eqref{inout-Z} gives a BTZ BH with non-zero  Euclidean sources. The inner product results
\begin{align}\label{innerprodBH}
\ln\langle\!\langle\Psi_F|\Psi_I\rangle\!\rangle\Big|_{BH} &\equiv \lim_{\Delta T\to0} i  S^0
=\int_\partial \phi_{E}(\tau',\varphi') \sum_{j\in\mathbb{Z}}\frac{(\Delta-1)^2}{2^{\Delta-1}\pi}\left[\cosh(\Delta \varphi+2\pi r_S j)-\cos(\Delta \tau)  \right]^{-\Delta}\phi_{E}(\tau,\varphi) \;,
\end{align}
where $\phi_{E}(\tau,\varphi)\equiv\phi_{I}(\tau,\varphi)\Theta(-\tau)+\phi_{F}(\tau,\varphi)\Theta(\tau)$, with $\Theta$ the step function. We have incorporated the appropriate two point function normalization according to \cite{Rastelli}.

In the low temperature limit, the result is
\begin{align}\label{innerprodThermal}
\ln\langle\!\langle\Psi_F|\Psi_I\rangle\!\rangle\Big|_{Th} &\equiv \lim_{\Delta T\to0} i  S^0
=\int_\partial \phi_{E}(\tau',\varphi') \sum_{j\in\mathbb{Z}}\frac{(\Delta-1)^2}{2^{\Delta-1}\pi}\left[\cosh(\Delta \tau+\beta j)-\cos(\Delta \varphi)  \right]^{-\Delta}\phi_{E}(\tau,\varphi) \;.
\end{align}
As expected from \cite{BTZ-Thermal}, the kernels in \eqref{innerprodBH} and \eqref{innerprodThermal} are connected via a double Wick rotation. 
By extending this computation for a complex field $\Phi_E$ and source $\phi_E$ we can identify these kernels as the K\"ahler potentials in the space of states, as observed in \cite{Aitor-Simplectic,Belin:2018bpg}.
In this sense \eqref{innerprodBH} and \eqref{innerprodThermal}, as well as the ones in \cite{us,us2}, provide explicit non-trivial examples.
Notice also that the states $|\Psi_{I,F}\rangle\!\rangle$ are not normalized nor orthogonal. 

Taking single derivatives of \eqref{Onchell} with respect to $\phi_{L/R}$ at high temperature, we get
\begin{align}\nn
\frac{\langle\!\langle\Psi_F|{\cal O}_L(t,\varphi)|\Psi_I\rangle\!\rangle}{\langle\!\langle\Psi_F|\Psi_I\rangle\!\rangle}\Big|_{BH}
&=\frac{(\Delta-1)}{2\pi^2 i r_S}\sum_l\int d\omega \;e^{-i\omega t+i l\varphi}\left( \bar\phi_F(-i\omega,l) e^{i\omega T/2}+\bar\phi_{I}(-i\omega,l)e^{-i\omega T/2}\right)\times \\
&
\qquad\qquad\qquad\qquad\qquad \qquad\qquad\qquad\qquad\qquad\qquad\times
\frac{\alpha_{\omega l\Delta}\beta_{\omega l\Delta}-\alpha_{-\omega l\Delta}\beta_{-\omega l\Delta }}{e^{2\pi\omega}-1}\label{O_L}
\end{align}

\begin{align}\nn
\frac{\langle\!\langle\Psi_F|{\cal O}_R(t,\varphi)|\Psi_I\rangle\!\rangle}
{\langle\!\langle\Psi_F|\Psi_I\rangle\!\rangle}\Big|_{BH}
&=\frac{(\Delta-1)}{2\pi^2 i r_S}\sum_l\int d\omega \; e^{-i\omega t+i l \varphi}  \left(e^{\pi\omega} \bar\phi_F(-i\omega,l) e^{i\omega T/2}+e^{-\pi\omega}\bar\phi_{I}(-i\omega,l) e^{-i\omega T/2}\right) \times \\
&\qquad\qquad\qquad\qquad\qquad \qquad\qquad\qquad\qquad\qquad\qquad\times
\frac{\alpha_{\omega l\Delta}\beta_{\omega l\Delta}-\alpha_{-\omega l\Delta}\beta_{-\omega l\Delta }}{e^{2\pi\omega}-1}\label{O_R}
\end{align}

We would like to highlight some aspects of these results. The $e^{\pm i\omega T/2}$ factors can be understood as arising from the distance between the location of the sources in the complex $t$-plane. Keep in mind that  ${\cal O}_{L/R}$ matrix elements are obtained by taking $T\to 0$ limit. A second observation is that $\phi_{I/F}$ contribute with both positive and negative frequencies, contrary to the zero temperature case \cite{us}. This is an expected  result related to the entanglement between L and R dofs and will be thoroughly discussed in the next section. Along these same lines, entanglement can readily be seen by noting that \eqref{O_L} and \eqref{O_R} differ by $e^{\pm \pi \omega}=e^{\pm \beta \omega/2}$ factors. 

In the low temperature regime the relevant geometry is depicted in Fig. \ref{Fig:ThermalAdS}. The result is
\begin{align}\label{ThermalO_L}
\frac{\delta S^0_{Th}}{\delta \phi_L}\Bigg|_{\phi_L=0}
=&\frac{\Delta-1}{2\pi^2}\sum_{nl} \left[\sum_\pm e^{\mp i\omega_{nl} t + i l\varphi} \left(\bar\phi_F(\mp i\omega_{nl},l)e^{i\omega_{nl} T/2} +\bar\phi_{I}(\mp i\omega_{nl},l)e^{-i\omega_{nl} T/2}\right)\right] \frac{s_{nl}(r)}{e^{\beta\omega_{nl}}-1}
\end{align}
\begin{align}\label{ThermalO_R}
\frac{\delta S^0_{Th}}{\delta \phi_R}\Bigg|_{\phi_R=0}
=&\frac{\Delta-1}{2\pi^2}\sum_{nl} \left[\sum_\pm e^{\mp i\omega_{nl} t + i l\varphi}  \left(e^{\mp\omega_{nl}(\sigma-\beta)}\bar\phi_F(\mp i\omega_{nl},l)e^{i\omega_{nl} T/2} +e^{\mp\omega_{nl} \sigma}\bar\phi_{I}(\mp i\omega_{nl},l)e^{-i\omega_{nl} T/2}\right)\right] \frac{s_{nl}(r)}{e^{\beta\omega_{nl}}-1}
\end{align}
where $\omega_{nl}$ and $s_{nl}$, defined in App. \ref{App:Thermal}, are the standard frequencies and normal modes in global AdS geometry and
\begin{equation}\nn
\bar\phi_{I}(- i\omega_{nl},l) \equiv \int_{\sigma-\beta}^0 d\tau d\varphi \;e^{\omega_{nl} \tau-il\varphi}\phi_{I}(\tau,\varphi) \;,
\qquad\qquad 
\bar\phi_{F}(- i\omega_{nl},l)\equiv\int^\sigma_{0} d\tau d\varphi \;e^{\omega_{nl} \tau-il\varphi}\phi_{F}(\tau,\varphi)\;.
\end{equation}
Notice that for general $\sigma$ the lhs of \eqref{ThermalO_L} and \eqref{ThermalO_R} do not refer to any bra/ket notation. This interpretation only appears at $\sigma=\beta/2$, where time reflection symmetry between the Euclidean pieces arises.

\section{Canonical quantization of the bulk fields and BDHM dictionary}
\label{Sec:BDHM}

In this section we will study the excited states \eqref{psi-rho-id} from the BDHM perspective \cite{BDHM}.  We will work in the large $N$ regime with free AdS fields in the probe limit. For completeness, we will review some relevant considerations on field quantization in BH geometries. Via the BDHM map, we will build the CFT ${\cal O}_R,{\cal O}_L $ operators from  quantized bulk fields. The coherent nature of the excited states will be demonstrated upon confronting with the outcome of the previous section. As a result, the excitations obtained by turning on sources in the Euclidean sections correspond to \textit{thermal} coherent states \cite{Thermal-Coherent}.

\subsection{Canonical quantization of scalar fields in a BH geometry}

Quantization of fields in a BH geometry gives rise to two sets of ladder operators. These can be seen to arise from the possibility of independent excitations in the L and R patches or, alternatively, from the two possible analytic extensions of the $L$-mode solutions to the $R$-patch when solving the problem in Einstein-Rosen coordinates. As well known, the corresponding vacuum states turn out to be non-equivalent. Here we follow the analytic approach. The outcome makes contact with the TFD dual theory discussed in Sec. \ref{Sec:ClosedPaths}.

We start quantizing the scalar field theory \eqref{scalarEOM} on the BTZ metric in Einstein-Rosen coordinates. Writing  $u^2=r^2-1$ \cite{ER}, the metric \eqref{metric-l2} turns into 
\begin{equation}
\label{metric-u}
ds^2 = -u^2 dt^2 + \frac{du^2}{u^2+1} + (u^2+1) d\varphi^2\,.
\end{equation}
The  $u\gtrless0$ regions correspond to the L and R patches respectively, and the $t$-coordinate coincides with that of the previous section. See Fig. \ref{Fig:Guille}(a) for a representation of the geometry\footnote{\label{foot}It is important to stress that $t$ has a boost-like character across the ER bridge. This can be seen when mapping the metric to Kruskal coordinates. Alternatively, one can resort to analyticity of the metric in the complex $u$ plane: for fixed $\Delta t>0$, proper time   $\Delta {\bf t}(u)=\int u \;d t>0$ at fixed $u>0$ flips sign when moving from  L ($u>0$) to  R ($u<0$). }. The KG field \eqref{scalarEOM} on the ER geometry is expanded as
\begin{align}\label{Qusol}
\hat \Phi(u,t,\varphi) = \sum_{l} \int_0^\infty d\omega \;  \hat d^{(1)}_{\omega l}\; h^{(1)}_{ \omega l}(u,t,\varphi) + \hat d^{(2)}_{ \omega l}\; h^{(2)}_{\omega l}(u,t,\varphi) + h.c.\;.
\end{align}
The positive energy modes $h^{(1,2)}_{ \omega l}$ are defined from the L and R modes which have support on their respective patches (aka Rindler-like modes). Explicitly, the L modes are defined as
\begin{equation}\label{Lmodes}
\left(\square -m^2\right)g_{L;\omega l}=0\; \qquad \partial_t g_{L;\omega l} = - i \omega g_{L;\omega l} \qquad  g_{L;\omega l} \equiv {\cal N}\!\!_{\omega l} \; e^{-i \omega t+i l \varphi} \left[f_{\omega l}(r)-f_{-\omega l}(r)\right], \qquad \omega>0
\end{equation} 
whereas the R modes are given by
\begin{equation}
g_{R;\omega l}\equiv g^*_{L;\omega l}. 
\end{equation}
The $u$-analytic $h$-modes are \cite{kenmoku,KeskiVakkuri}
\begin{equation}\label{hmodes}
h^{(1)}_{\omega l}(u,t,\varphi) = \frac{1}{\sqrt{2 \sinh(\pi \omega)}} \begin{cases} e^{\pi \omega/2} \; g_{L;\omega l} & \text{on L}\\ e^{-\pi \omega/2}\; g^*_{R;\omega l} & \text{on R}\end{cases} 
\qquad\qquad 
h^{(2)}_{\omega l}(u,t,\varphi) = \frac{1}{\sqrt{2 \sinh(\pi \omega)}} \begin{cases} e^{-\pi \omega/2} \; g^*_{L;\omega l} & \text{on L}\\ e^{\pi \omega/2} \; g_{R;\omega l} & \text{on R}\end{cases}
\end{equation}
c-numbers ${\cal N}\!\!_{\omega l}$ guarantee that $g$-modes are orthonormal on their respective Rindler-like patches\footnote{The scalar product is defined in  standard fashion 
$$
\langle\phi_1,\phi_2\rangle=-i\int_\Sigma(\phi_1\partial_\mu\phi_2^*-\phi_2^*\partial_\mu\phi_1)\,n^\mu\sqrt\gamma dud\varphi\,, $$
with $n^\mu$ the unit normal to the constant $t$-hypersurface $\Sigma$ and  $\gamma_{ij}$ its induced metric.  
A sign change in $n^\mu$ between L and R patches ($u\gtrless0$) arises from the boost character of $t$ depicted in Fig.\ref{Fig:Guille}(a).}. The vacuum state defined as
\begin{align}
\hat d^{(1)}_{\omega l}|\Psi_0\rangle\!\rangle  = 
\hat d^{(2)}_{\omega l}|\Psi_0\rangle\!\rangle  = 0\;,\label{TFDvac}
\end{align}
corresponds to the so-called TFD vacuum state defined below \eqref{unit}.

The $h$-modes presented in this section are Unruh-like, we have built them from: ER coordinates, which cover the exterior of the BH, and by demanding analyticity in the radial coordinate across the wormhole. Our viewpoint aims at studying CFT information attainable holographically from the BH exterior.

\subsection{BDHM at finite temperature, TFD doubling and coherence}

In this section we review the BDHM dictionary \cite{BDHM} at finite temperature. The standard prescription defines  quantum local CFT operators $\hat {\cal O}(t,\varphi)$ from AdS quantized fields $\hat \Phi (r,t,\varphi)$ via the map
\begin{equation}
\hat {\cal O}(t,\varphi)\equiv 2(\Delta-d) \lim_{r\to\infty}r^{\Delta}\; \hat{\Phi}(r,t,\varphi)\;,
\end{equation}
where $r\to\infty$ defines the unique asymptotic boundary of the bulk theory at zero temperature. The coordinate dependent $r^{\Delta}$ factor conspires to give a finite limit and the $2(\Delta-d)$ factor is required to have a precise matching with the GKPW prescription results \cite{BDHM2, us}. 
At finite temperature, in agreement with the TFD approach described in Sec. \ref{Sec:ClosedPaths}, dof's get duplicated in the gravity theory, this manifests in two disconnected asymptotic boundaries. Hence, from \eqref{Qusol} we define
\begin{align}
\hat {\cal O}_L(t,\varphi)\equiv (2\Delta-d)\lim_{u\to\infty} \hat \Phi(u,t,\varphi) = \sum_{l} \int_0^\infty d\omega \; \hat d^{(1)}_{\omega l}\; e^{\pi \omega/2} e^{-i \omega t+i l \varphi} {\cal O}_{\omega l}  + \hat d^{(2)}_{\omega l}\;e^{-\pi \omega/2} e^{+i \omega t-i l \varphi} {\cal O}^*_{\omega l} + h.c.\;, \label{OL}\\
\hat {\cal O}_R(t,\varphi)\equiv (2\Delta-d)\lim_{u\to-\infty} \hat \Phi(u,t,\varphi) = \sum_{l} \int_0^\infty d\omega \; \hat d^{(1)}_{\omega l}\; e^{-\pi \omega/2} e^{-i \omega t+i l \varphi} {\cal O}_{\omega l}  + \hat d^{(2)}_{\omega l}\;e^{\pi \omega/2} e^{+i \omega t-i l \varphi} {\cal O}^*_{\omega l} + h.c.\;,\label{OR}
\end{align}
where the c-numbers
\begin{equation}\label{Odef}
{\cal O}_{\omega l} \equiv \frac{(2\Delta-d)}{\sqrt{2 \sinh(\pi \omega)}}   \; {\cal N}\!\!_{\omega l} \left[ \alpha_{\omega l}\beta_{\omega l}-\alpha_{-\omega l}\beta_{-\omega l} \right]\;,
\end{equation}
are  inherited from the modes normalization. 

The excited state \eqref{psi-rho-id} in the Interaction Picture, built from \eqref{OR} becomes\footnote{As can be seen from \eqref{TFDvac}, 
the actions of ${\cal O}_L$ and ${\cal O}_R$ on the TFD vacuum are related. As a consequence, one can pick any of them to build \eqref{psi-rho-id} and  the excitations over $|\Psi_0\rangle\!\rangle$ are physically equivalent. This issue will be further elucidated in the upcoming section.}
\begin{equation}\label{initial-state}
|\Psi_I\rangle\!\rangle \equiv {\cal P}\left\{ e^{-\int_{-\pi}^0 d\tau \;\hat {\cal O}_R(\tau) \phi_I(\tau) } \right\}|\Psi_0\rangle\!\rangle \propto \exp\left\{\sum_l \int_0^\infty d\omega\; \lambda_{I;\omega l}^{(1)}  \hat d^{(1) \dagger}_{\omega l} + \lambda_{I;\omega l}^{(2)} \hat d^{(2) \dagger}_{\omega l} \right\}|\Psi_0\rangle\!\rangle\,.
\end{equation}
where
\begin{equation}
\label{eigen-v}
\lambda_{I;\omega l}^{(1)} =- e^{-\omega \pi/2} \bar \phi_I(-i\omega,l) \;{\cal O}_{\omega l}^*
\qquad\qquad
\lambda_{I;\omega l}^{(2)} =-  e^{\omega \pi/2} \bar \phi_I(+i\omega,l) \;{\cal O}_{\omega l} \;.
\end{equation}
In obtaining these expressions we exploited  standard disentangling theorems \cite{Gilmore-Squeezed}.
Notice the similarity of the rhs of \eqref{initial-state} with the zero temperature expression \eqref{coherent-state}. Notice that the operator multiplying $|\Psi_0 \rangle\!\rangle$ on the r.h.s. of \eqref{initial-state} is nothing but the form  of the (sourced) evolution operator \eqref{Upm}, represented in the Interaction Picture in the bulk field theory, and by virtue of \eqref{TFDvac}, this can then be rewritten as a unitary \emph{displacement} operator up to a constant factor. 

Results \eqref{initial-state} and \eqref{eigen-v}, which are the main result of this work,  demonstrate that the states obtained by turning on sources in the Euclidean sections are {\it thermal} coherent states. We conclude that the coherent/ semi-classical character of the states \eqref{psi-rho-id}, originally developed at zero temperature in \cite{us,us2}, remains valid at finite temperature, as claimed in \cite{prev}. In order to put this result into the more familiar finite temperature language \cite{Thermal-Coherent}, the corresponding (unnormalized) density matrix for the excited states \eqref{psi-rho-id} using \eqref{Upm} is,
\begin{equation}\nn
\rho_\phi \equiv \text{Tr}_{\tilde{{\cal H}}}\; \, |\Psi_\phi \rangle\!\rangle \langle \! \langle \Psi_\phi | \,=\, U^{}_\phi \; U_\phi^\dagger\;
\end{equation}
that is manifestly hermitian (see \cite{prev}). Thus, by interpreting this expression in the bulk (free-field) theory, one can express it in the Interaction Picture as $\rho_\phi \, \propto\, U_\phi \, e^{-\beta H} \, U_\phi^\dagger$, where $U_\phi$ is a displacement operator, consistently with \eqref{initial-state}. This is the standard density matrix of a thermal coherent state.

Direct computation using \eqref{OL}, \eqref{OR} and \eqref{initial-state} leads to 
\begin{align}
\frac{\langle\!\langle\Psi_F| \hat {\cal O}_L(t,\varphi) |\Psi_I\rangle\!\rangle }{\langle\!\langle\Psi_F|\Psi_I\rangle\!\rangle}
&= - \sum_{l} \int_{-\infty}^\infty d\omega \;e^{-i \omega t+i l \varphi} \left( \bar \phi_F(-i\omega,l) +\bar \phi_I(-i\omega,l) \right)  |{\cal O}_{\omega l}|^2 \label{OL1p} 
\end{align}
\begin{align}
\frac{\langle\!\langle\Psi_F| \hat {\cal O}_R(t,\varphi) |\Psi_I\rangle\!\rangle }{\langle\!\langle\Psi_F|\Psi_I\rangle\!\rangle}
&= - \sum_{l} \int_{-\infty}^\infty d\omega \; e^{-i \omega t+i l \varphi}  \left( e^{\pi \omega} \bar \phi_F(-i\omega,l) + e^{-\pi \omega}\bar \phi_I(-i\omega,l)\right) |{\cal O}_{\omega l}|^2  \label{OR1p} 
\end{align}
where one can immediately recognize the \eqref{O_L} and \eqref{O_R} structure among the sources. 
Further comparison of these expressions provides an analytic expression for $|{\cal O}_{\omega l}|^2$, determining in turn ${\cal N}\!\!_{\omega l}$, which to the author's  knowledge is not present in the literature. See \cite{kenmoku} for an integral expression of ${\cal N}\!\!_{\omega l}$ where the need for a careful computation involving regulators is emphasized. One can check that in the zero temperature limit, as  in the Thermal-AdS regime presented here, where the normalization factors are known, the corresponding expressions exactly match \cite{us}. Similar observations were made recently in \cite{Iqbal}\footnote{We thank Ra\'ul Arias for pointing out this reference to us.}. The precise expression for the eigenvalues \eqref{eigen-v} is obtained by comparing the path integral and BDHM approaches. 

\subsection{On the Unruh's Trick in the TFD formulation}

This section is devoted to show how Unruh's trick \cite{Unruh,carroll}, manifestly realized in our geometry through identical Euclidean sections, plays a crucial role in the field quantization by defining the vacuum state through a quantum constraint and, moreover, how it naturally generalizes for excited states.
This could also be seen as an alternative formulation of the Unruh-Hawking effect.

First, recall that in the TFD context $\hat \Phi_R$ and $\hat \Phi_L$, the quantized counterparts of \eqref{Rsol} and \eqref{Lsol2}, map into each other by the tilde conjugation rules \eqref{Tilde-rules}, and although they are independent dof's, their action on the TFD vacuum state is not. In fact, the TFD vacuum is completely determined by the constraint,
\begin{equation}\label{TFDconstraint}
\left[\hat \Phi_L(|u|, t=T_I, \varphi) - \hat \Phi_R(-|u|, t=T_I-i\pi, \varphi)\right]|\Psi_0\rangle\!\rangle\;=0\;,\qquad\qquad \forall u, \varphi\;,
\end{equation}
complemented with a similar equation for the canonically conjugated momentum fields.
The physical meaning of this is that the vacuum state, whose wave functional is described by the Euclidean geometry of Fig. \ref{Fig:Guille}(b), is such that acting with an $L$-operator on it at $T_I$, is the same as acting with the $R$-operator at $T_I$ but evolved  $-i\pi$ in imaginary time.

Inserting \eqref{Qusol} and \eqref{hmodes} above and using orthonormality of the modes one gets 
\begin{align}
\hat d^{(1)}_{\omega l}|\Psi_0\rangle\!\rangle \propto \left(\hat L_{\omega l}-e^{-\omega\beta/2} \hat R_{\omega l}^\dagger \right)|\Psi_0\rangle\!\rangle =0 
\qquad
\hat d^{(2)}_{\omega l}|\Psi_0\rangle\!\rangle \propto \left(\hat L_{\omega l}^\dagger-e^{+\omega\beta/2} \hat R_{\omega l}\right)|\Psi_0\rangle\!\rangle = 0\;,\label{constraint0}
\end{align}
where $\hat L^\dagger_{\omega l}$ and $\hat R^\dagger_{\omega l}$ create excitations with support on L and R wedges respectively, and $\hat{d}^{(1,2)}_{\omega l}$ are defined so that they anihilate the state $|\Psi_0\rangle\!\rangle$. These relations are known as thermal state conditions and define the Bogoliubov transformation between both sets of ladder operators\footnote{Similar formulations in the string context can be found in Refs. \cite{Abdalla-StringTFD,Boots-StringTFD}}. Notice that in \eqref{constraint0} we have reintroduced the explicit dependence on the temperature $\beta$.

An important consequence of this formulation is that the modes associated to operators $\hat{d}^{(1,2)}_{\omega l}$ are precisely the linear combinations \eqref{hmodes} of the $L,R$ solutions, which are analytic through the throat $u=0$. This captures the features discussed in Sec. \ref{Sec:Analitycity}. It is also worth noticing that this is consistent with the fact that the points on the throat of the wormhole, i.e. $u=0$, are fixed points of the evolution operator $U_0(\tau)$ of the bulk quantum theory, whose Hamiltonian is the boost generator, and its analytic extension to imaginary times evolves the hipersurfaces depicted as red lines in Fig \ref{Fig:Guille}(b). In this sense, $|\Psi_0\rangle\!\rangle$ is the thermal KMS state with respect to the generator of the Lorentz boosts \cite{Bisognano}.

A novel remarkable fact is that by performing an imaginary $-i\pi$ time translation with the sourced evolution operator \eqref{Upm} in place of $U_0$, one gets
\begin{equation}\nn
\hat \Phi_R(T_I-i\pi) \equiv U_\phi(\pi)\,\, \hat \Phi_R(T_I) \,\,U^\dagger_\phi(\pi)\,\,.
\end{equation} 
Recall that in this formulation all the fields are represented in the Interaction Picture.
This can be used to define an (initial) excited state since the constraint (\ref{TFDconstraint}) now becomes
\begin{equation}\label{TFDconstraint-phi}
\left[\hat \Phi_L(|u|, t=T_I, \varphi)\,-\, \,U_\phi(\pi) \,\hat \Phi_R(-|u|, t=T_I, \varphi) \, U^\dagger_\phi(\pi) \,\right] |\Psi_\phi\rangle\!\rangle = 0 \;,\qquad\qquad \forall u, \varphi\;,
\end{equation}
complemented also with the corresponding equation for the canonically conjugated momentum. The frequency decomposition of these equations now yields
\begin{equation}\label{constraint-phi}
\left ( \,\hat{d}^{(1,2)}_{\omega,l} - \lambda^{(1,2)}_{\omega,l} \,\right)\, |\Psi_\phi\,\rangle\!\rangle=0
\end{equation}
where we have used that the (adequately normalized) operator $U_\phi$ acts on ladder operators of the bulk theory as a displacement, i.e.
$$U_\phi(\pi) \, \hat{d}^{(1,2)\,}_{\omega l} U^\dagger_\phi(\pi)
= \hat{d}^{(1,2)}_{\omega l} + \lambda^{(1,2)}_{\omega l}\;$$ 
according to the arguments below \eqref{initial-state}.

Thus, the solution of \eqref{constraint-phi} is clearly the state \eqref{initial-state}.
This presents an alternative perspective on our prescription of Sec. \ref{Sec:ClosedPaths} for the excited initial/final states in the boundary field theory.

\section{Discussion and Conclusions}
\label{Sec:Conclusion}

In a previous article \cite{prev}, we presented the gravity dual of a finite temperature real time CFT, casted in TFD formulation, and computed real time two-point functions. The Schwinger-Keldysh path on which the CFT was defined has two possible dual geometries: a real time extension of Thermal-AdS and a novel geometry consisting of glued Euclidean and Lorentzian AdS-BH pieces, see Fig. \ref{Fig:Camino}(b), which dominate below and over the critical Hawking-Page temperature respectively. Both geometries contain two equal length Euclidean pieces that within the TFD formalism are naturally associated with initial and final states of the system. The geometries also contain two causally disconnected Lorentzian regions, L and R, which correspond to the two TFD copies of the system.

In this work, following ideas in \cite{us,prev}, we have studied holographic excited states by turning on asymptotic sources on the Euclidean regions. The resulting states were shown to be coherent states wrt the TFD vacuum, i.e. the excitations are not described in terms of either the L or R dofs but rather Bogoliubov transformed counterparts. Such excitations are known as \textit{thermal} coherent states. A precise expression for the eigenvalues was also given in terms of the Euclidean sources profile. These results extend  the work in \cite{us} to the case of finite temperature systems. 
We stress that our main objective here was to characterize these holographic excited states and not the mixed signature manifold, which was merely a device to study their properties. For example, we can now consider them as initial states to study evolution of information in geometries that consider the BH interior as in the toy model considered in \cite{eternal}.

Our study revealed itself to be particularly interesting in the high temperature limit where bulk real time regions L and R get connected though an ER wormhole. For this geometry, we found that the  analyticity of the field through the  spacelike gluing surfaces, imposed by construction, extends to the radial coordinate connecting the L and R regions across the wormhole. As a consequence, the field solution to the equations of motion on the mixed signature geometry encapsulates Bogoliubov coefficients between the L and R degrees of freedom. In this sense, our bulk can be interpreted as the geometrical embodiment of the standard Unruh trick. For the BDHM approach, Unruh-like modes were built solely from the exterior of the BH and analyticity on the radial ER coordinate across the wormhole. This is an interesting result for the study of CFT information accessible from the outside of event horizons.

Two objects were of interest in characterizing our states:  inner products and  matrix elements of local single trace operators. The latter, computed with two equivalent prescriptions, was key to determine the eigenvalues of the coherent states. From a path integral formulation a complete field solution with sources turned on was built, which required the study of N and NN modes in half Euclidean BTZ geometries. BDHM approach requires  positive energy eigenstates normalization factors, for which we didn't find an analytic expression in the literature. Thus, by comparing the path integral and BDHM results we were able to give a closed expression for the eigenvalues and the normalization factors. The inner product, on the other hand, can be thought as a reinterpretation of the free energy of the geometry configuration with sources turned on. On that regard, one expects on general grounds that our results should be valid away from the Hawking-Page critical temperature. The kernels on the inner product of these holographic excited states for complex bulk fields has been recently recognized in \cite{Aitor-Simplectic,Belin:2018bpg} as the Kähler potential in the space of states. Though presented for real scalars, our results here as well as in \cite{us,us2} are immediately extended to the complex scenario.

A number of trails open up for future work. We are currently working on holographic computations of relative and entanglement entropies using these holographic excited states. We leave generalizations of our geometry to future works: Euclidean pieces with unequal lengths might be related to some type of back-reaction, and may play a role in the study of traversable wormholes \cite{Raulo,Pingao} and out of equilibrium systems \cite{BootsBrazil}. We also plan to consider special SK paths in order to compute OTOC's \cite{OTOC} relevant to  the chaos context. Related to this, it would be interesting to understand the role of multiple Euclidean pieces in relation to the analyticity of the fields through the wormhole and whether the TFD interpretation still holds. Further studies on the nature of these holographic excited states are also already under development. We would also like to explore our coherent states as as a generating base of the complete Fock space, which should require careful backreaction treatment.

~

\nin {\bf Acknowledgements}

\nin  Work supported by UNLP and CONICET grants X791, PIP 2017-1109 and PUE B\'usqueda de nueva F\'\i sica.

\newpage

\appendix

\section{Low temperature excited states: Thermal AdS}\label{App:Thermal}

\begin{figure}[t]\centering
\begin{subfigure}{0.49\textwidth}\centering
\includegraphics[width=.9\linewidth] {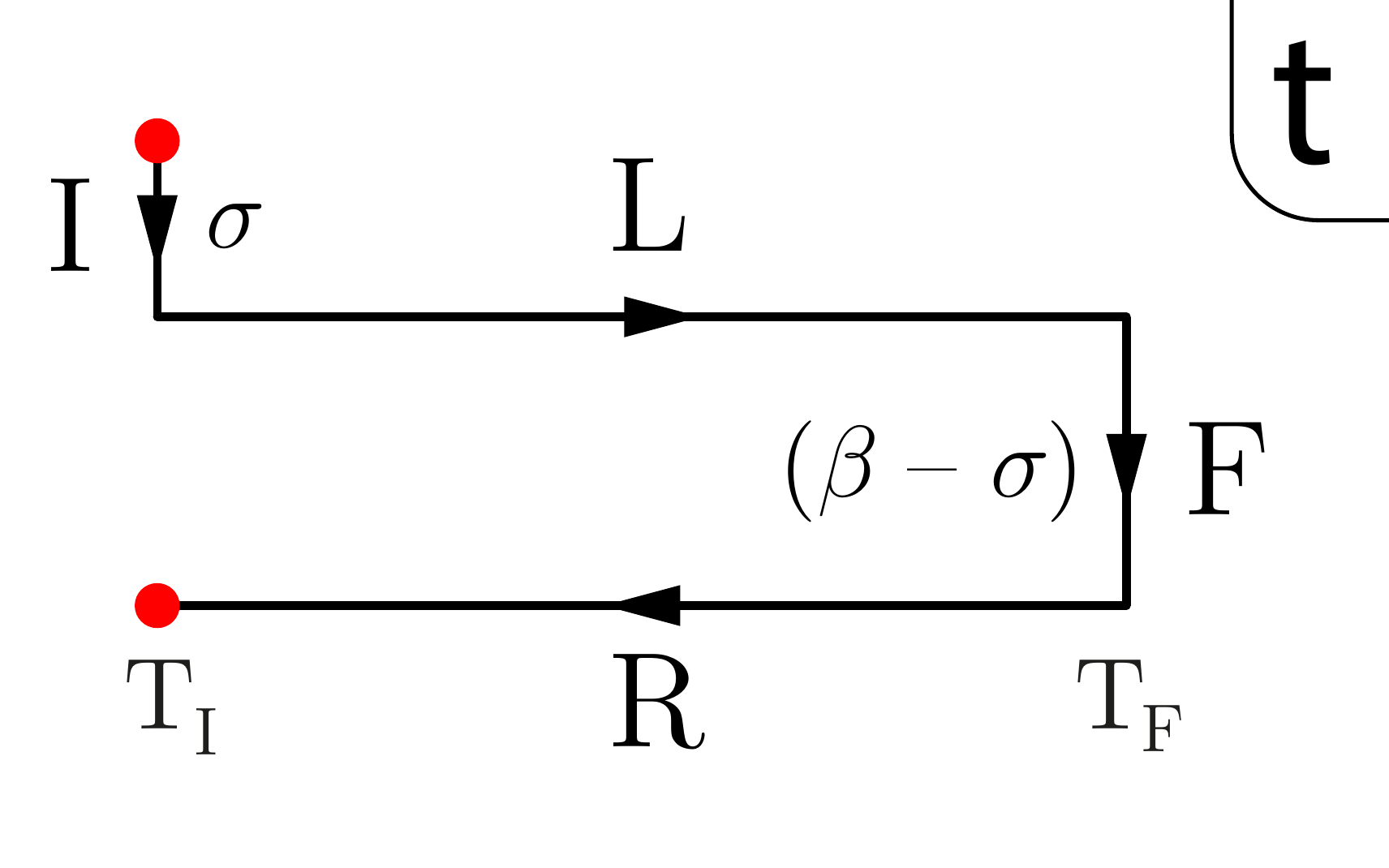}
\caption{}
\end{subfigure}
\begin{subfigure}{0.49\textwidth}\centering
\includegraphics[width=.9\linewidth] {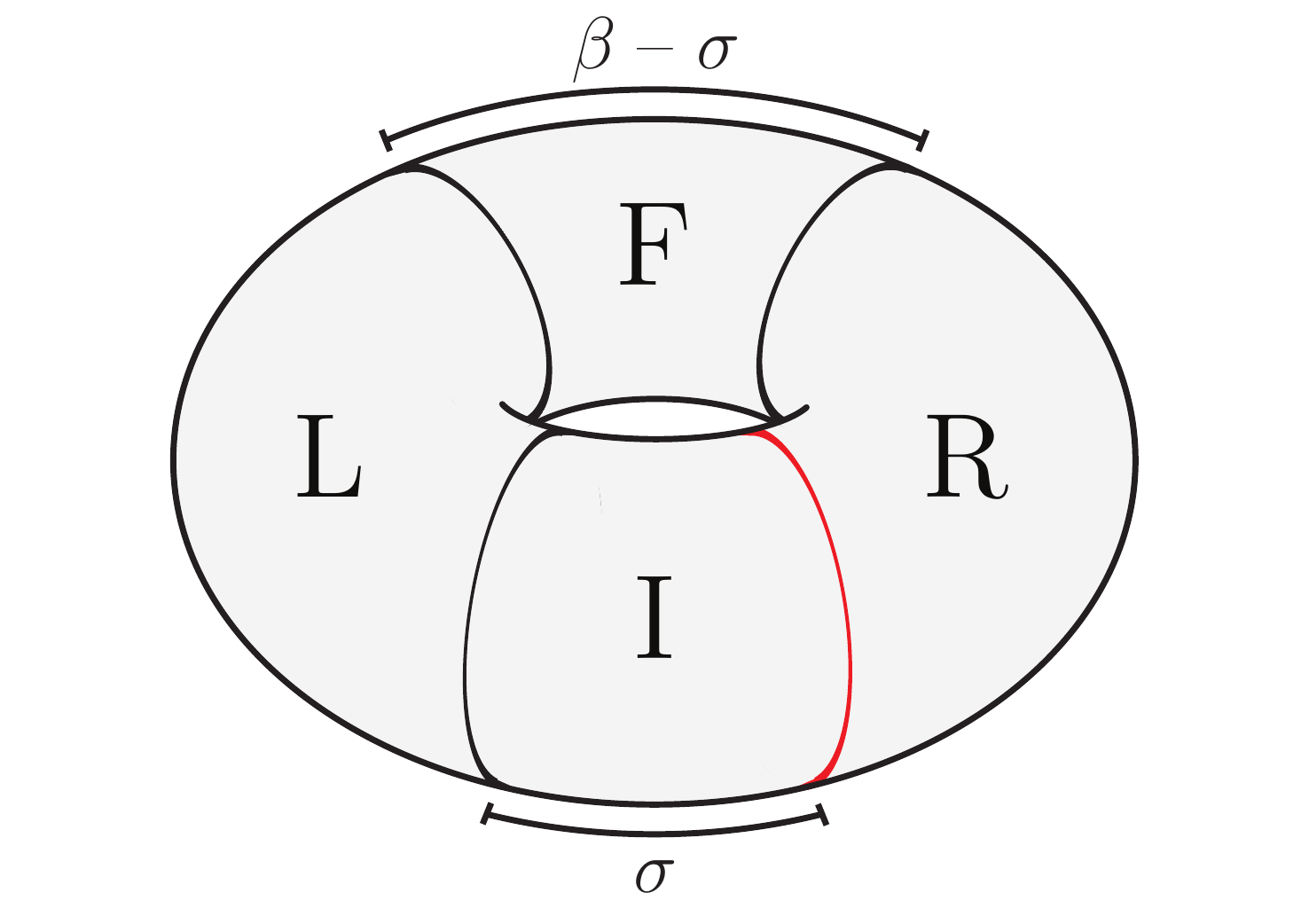}
\caption{}
\end{subfigure}
\caption{(a) Variation on the closed Schwinger-Keldysh path of Fig. \ref{Fig:Camino}(a) with Euclidean segments of lengths $\sigma$ and $\beta-\sigma$  (b) Real time extension of the Thermal-AdS geometry dual of the path. Contrary to Fig. \ref{Fig:Camino}(b), the Lorentzian pieces are disconnected and thus admit arbitrary lengths of the Euclidean pieces.}
\label{Fig:ThermalAdS}
\end{figure}

In this Appendix we summarize the computations in the low temperature geometry dual to the path in Fig. \ref{Fig:Camino}(a). This geometry is built by inserting two Lorentzian pure AdS segments in the standard Euclidean Thermal AdS geometry. These Lorentzian segments however evolve in opposite directions. This construction results in Fig. \ref{Fig:ThermalAdS}. The Lorentzian sections L and R are still entangled in this regime though not topologically connected. As a consequence, analyticity inside the bulk does not restrict the lengths of the Euclidean sections which we take to be $\beta-\sigma$ and $\sigma$ respectively for I and F. The $\sigma=\beta/2$ path is still preferred as it recovers the natural map between vectors in the Hilbert space and its dual. From the CFT point of view, this is a well known result \cite{UmezawaAFT} that privileges the TFD interpretation/framework above other SK paths.

One needs to build a field solution for general sources and obtain inner product and matrix elements. The computations in this geometry are mostly direct from the zero temperature case.
For the Lorentzian regions, we have the action and EOM \eqref{scalarEOM} over the pure AdS$_3$ metric
\begin{equation}\nn
ds^2=-(r^2+1)dt^2+\frac{dr^2}{r^2+1}+r^2d\varphi^2\;.
\end{equation}
A plane wave expansion $\Phi\propto e^{-i \omega t + i l \varphi} s(\omega,l,r)$ leads to a differential equation for $s(\omega,l,r)$. Regularity in the bulk fixes
\begin{equation}\nn
s(\omega,l,r) = \frac{\Gamma \left(\frac{1}{2} (|l|+\Delta -\omega )\right) \Gamma \left(\frac{1}{2} (|l|+\Delta +\omega )\right)}{\Gamma (\Delta -1) \Gamma (|l|+1)} \left(1+r^{2}\right)^{\omega/2}\,\, r^{|l|}\,\,
_2F_1\left( \frac{\omega+|l|+\Delta}{2}, \frac{\omega+|l|-\Delta+2}{2} ; 1+|l| ; -r^{2}\right)\,,
\end{equation}
with the overall constant fixed so $s(\omega,l,r) \sim r^{\Delta}+\dots$ for generic $\{\omega,l\}$. This normalization puts singularities on the real $\omega$ axis.
The residues of these poles $\omega_{nl}=2n+\Delta+|l|$ can be used to define the N modes
\begin{equation}
\nn
s_{nl}(r)\equiv \oint_{\omega=-\omega_{nl}}d\omega \;s(\omega,l,r) \,,\qquad \omega_{nl}=2n+\Delta+|l|\;.
\end{equation}

The most general solution on L is 
\begin{align}
\Phi_L(r,t,\varphi)= &\frac{1}{4\pi^2 } \sum_{l\in\mathbb{Z}} \int_{\mathcal{F}} d\omega e^{-i \omega t + i l \varphi}  \bar\phi_L(\omega,l) s(\omega,l,r) 
+ \sum_{\substack{n\in\mathbb{N}\\ l\in\mathbb{Z} }}\Big(
 L_{nl}^{+}\, e^{- i \omega_{nl} t}+ L_{nl}^{-}\, e^{+ i \omega_{nl} t}\Big)e^{  i l \varphi}  s_{nl}(r)\,,
 \nn
\end{align}
where the Feynman path $\cal F$ was chosen in the first term and the arbitrary coefficients $L_{nl}^{\pm}$ will be determined once we impose boundary conditions \eqref{bc}. An analogous expression for R can be written and we explicitly present the Euclidean solution in I,
\begin{align}
\Phi_I(r,\tau,\varphi)= &\frac{1}{4\pi^2 } \sum_{l\in\mathbb{Z}} \int d\omega  e^{i \omega \tau + i l \varphi}  \bar \phi_I(-\omega,l) s(-i\omega,l,r)
+ \sum_{\substack{n\in\mathbb{N}\\ l\in\mathbb{Z} }}
 \left( I_{nl}^{+}\, e^{-  \omega_{nl} \tau }+I_{nl}^{-}\, e^{  \omega_{nl} \tau } \right) e^{ i l \varphi} s_{nl}(r)\,,
\nn
\end{align}
to fix notation and conventions. Notice that the subindex $\cal F$ is no longer required as the poles of $s(-i\omega,l,r)$ lie away from the real axis.
Following analogous steps as in \cite{us}, one can use the gluing conditions \eqref{bc} to uniquely fix the coefficients $L^{\pm}_{nl}$, $I^{\pm}_{nl}$ as well as their R and II counterparts. The computations are more tedious than pedagogical, one essentially reduces the problem to a set of lineal equations with a unique solution for the coefficients. 
As an example, we present the coefficients $I_{nl}^\pm, \;F_{nl}^\pm, \;R_{nl}^\pm$ due to a source on L which is related to the eigenvalues of the initial excited states on the ${\cal O}_L$ operator:
\begin{equation}\nn
L^{\pm}_{nl}=\frac{1}{4\pi^2} \frac{\phi_L(\pm \omega_{nl},l)}{e^{\omega_{nl} \beta}-1} \qquad I_{nl}^\pm = e^{\pm i \omega T/2} L^{\pm}_{nl} \qquad R_{nl}^\pm = e^{\mp\omega \sigma} L^{\pm}_{nl} \qquad F_{nl}^\pm = e^{\mp i \omega T/2} L^{\pm}_{nl}\;,
\end{equation}
which are to be compared to \eqref{coeffsL}. Solving for a source on every region leads to the complete on shell action and from there get the inner product and matrix elements, shown in \eqref{innerprodThermal}, \eqref{ThermalO_L} and \eqref{ThermalO_R}.

Regarding the BDHM computations carried on in Sec. \ref{Sec:BDHM} for the BH, the analogous Thermal scenario is much simpler and less rich. The throat is absent and the real time theories are entirely disconnected if not through the Euclidean regions, cf. Figs. \ref{Fig:Camino}(b) and \ref{Fig:ThermalAdS}(b). Each Lorentzian segment has an independent quantization which is the standard zero temperature computation carried in \cite{us}. Only the $\sigma=\beta/2$ path leads to sensible Hermitian conjugation rules \cite{Umezawa:TFD=beta/2,UmezawaAFT}, otherwise ad-hoc factors must be added to successfully go back and forth. In this set-up, the excited state mimics the structure of \eqref{initial-state}, but the basis is discreet in Global AdS coordinates,
\begin{equation}\nn
|\Psi_I\rangle\!\rangle \equiv {\cal P}\left\{ e^{-\int_{-\pi}^0 d\tau \;{\cal O}_R(\tau) \phi_I(\tau) } \right\}|\Psi_0\rangle\!\rangle \propto \exp\left\{\sum_{nl}\;\lambda_{I;nl}^{(1)}  d^{(1) \dagger}_{nl} +\lambda_{I;nl}^{(2)}  d^{(2) \dagger}_{nl}\right\}|\Psi_0\rangle\!\rangle
\end{equation}
where $d^{(i) \dagger}$, $i=1,2$ combines positive and negative energy excitations of the L and R regions mixed by the standard Bogoliubov transformation,
\begin{equation}\nn
\lambda_{I;nl}^{(1)} =-e^{-\omega_{nl} \pi/2} \bar \phi_I(-i\omega_{n l},l) \;{\cal O}_{n l}^*
\qquad\qquad
\lambda_{I;nl}^{(2)} =-e^{\omega_{nl} \pi/2} \bar \phi_I(+i\omega_{n l},l) \;{\cal O}_{n l} \;.
\end{equation}
and ${\cal O}_{n l}$ the zero temperature inherited operator coefficients discussed in \cite{us}.

\end{document}